\documentclass[12pt,draftclsnofoot,journal,a4paper,oneside,onecolumn]{IEEEtran}
\usepackage{latexsym}
\usepackage{graphicx}
\usepackage{amsfonts,amssymb,amsmath}
\usepackage{hyperref}
\usepackage{subfigure}
\usepackage{cite}
\usepackage{color}
\newtheorem{theorem}{Theorem}
\newtheorem{proposition}{Proposition}

\def\BibTeX{{\rm B\kern-.05em{\sc i\kern-.025em b}\kern-.08em T\kern-.1667em\lower.7ex\hbox{E}\kern-.125emX}}
\begin{document}

\title{Performance Characterization of Relaying Using Backscatter Devices}

\author{Xiaolun~Jia,~\IEEEmembership{Student Member,~IEEE,}
        and~Xiangyun~Zhou,~\IEEEmembership{Senior Member,~IEEE}
\thanks{This work was supported by the Australian Research Council's Discovery Project Funding Scheme under Project DP170100939. Part of this work was presented in \cite{gc}.}
\thanks{The authors are with the Research School Electrical, Energy and Materials of Engineering, The Australian National University, Canberra, Australia (email: \{xiaolun.jia, xiangyun.zhou\}@anu.edu.au).}}

\maketitle
\begin{abstract}
In this paper, we examine the error performance of backscatter communication in the presence of ambient interference, where the backscatter device acts as a relay. Specifically, the performance comparison of amplify-and-forward (AF) and decode-and-forward (DF) backscatter relaying is considered for the first time. Considering energy-based detection for on-off keying (OOK) modulation, we derive the statistics of the received signal power, from which the detection thresholds and corresponding bit error rates (BER) are obtained analytically. For the DF scheme, we allow the source node to transmit continuous-wave signals during the relay-to-destination transmission phase to power the backscatter relay. Under a total power budget constraint at the source, we optimize the power allocation for the transmissions in the source-to-relay and relay-to-destination phases. Numerical analysis shows that the DF scheme with optimal power allocation performs similarly compared to the AF scheme, despite the added complexity of the decoding operation. On the other hand, the AF scheme significantly outperforms the DF scheme when the reflection coefficients at the backscatter device do not correspond to perfect OOK. These results provide valuable insights into the design and deployment of backscatter nodes with the goal of improving coverage.
\end{abstract}


\section{Introduction}

\subsection{Background and Motivation}
\IEEEPARstart{T}{he} proliferation of the Internet of Things (IoT) paradigm in recent years has invariably resulted in the deployment of massive numbers of low-power sensors, which monitor and gather data from the surrounding environment. These devices have lower complexity than conventional cellular user equipment (UE) and can communicate with central entities and each other autonomously. Nonetheless, such devices perform active transmissions, which incur significant power consumption. Batteries, which are the preferred mode of power supply in these devices, are prone to being quickly exhausted. As future networks are expected to be comprised of billions of such devices, the amount of effort required to replace batteries could quickly become infeasible.

Backscatter communication has received increased research attention in recent years as a way for low-complexity devices to ease their reliance on battery power. The concept of communication by reflection of radiofrequency (RF) signals has been extensively applied to radiofrequency identification (RFID) systems. Low-power transceivers, referred to as tags, are powered by a continuous wave (CW) signal originating from a reader. Each tag modulates data onto the CW signal by switching its antenna between different load impedances corresponding to reflecting states, with the information-bearing carrier signal then returning to the reader. Previous studies in \cite{kimionis2013bistatic, liu2013ambient, parks2015turbocharging, bharadia2015backfi, wang2017fm} demonstrated the practicality of backscatter systems from both theoretical and implementation perspectives, with emphasis on interoperability with both unmodulated and modulated ambient signals. Using this mechanism, the lifetimes of such devices could be significantly extended, as active transmissions are not required.

Much research effort has been devoted to improving the reliability of backscatter communication in terms of detection performance, range and coverage. Complete link budget expressions of the monostatic architecture were presented in \cite{GD09}, where the maximum range was shown to be up to several meters. The design of optimal backscatter reflection coefficients was considered in \cite{improving}; while the authors in \cite{QAM12} examined the use of higher-order modulation and coding to improve range and spectral efficiency. For the bistatic architecture, where the RF source is separated from the reader, detection performance was studied under both coherent and noncoherent cases \cite{bistatic, CohFSK15, NCBi17}, to realize the potential of an order of magnitude increase in range. Moreover, the performance of ambient backscatter systems were also characterized in works such as \cite{Amb16, noncoherent, qian2017semi, exactBER}. From a coverage perspective, the work in \cite{HH17} and \cite{BC17} examined the network throughput and outage probability using stochastic geometry; although the analysis was limited by the achievable range of monostatic systems.

A second, less-studied use case for backscatter devices is their use as relays, rather than data sources. In light of the joint requirements of IoT networks on coverage and low power consumption while maintaining acceptable detection performance, relay by backscatter appears to be an efficient alternative to powered relays. Backscatter relays differ from conventional relays due to their passive modulation and demodulation, instead of using power amplifiers for transmission and performing complex signal processing operations to recover a signal. Given the speedy advancement in the capability of backscatter devices, it is not unreasonable to predict that they may achieve similar levels of coverage compared to conventional relays in the near future, with far less need for maintenance.

Work in \cite{BFRelay} was among the first to explore the backscatter relay use case, where a base-station-aided relaying protocol was considered to enable communication between two distant backscatter devices. The operation of backscatter relays in \cite{BFRelay} was similar to the traditional amplify-and-forward (AF) protocol, in that each node selected a load impedance and reflected incoming signals without processing. Work in \cite{Mun19} considered a similar base-station-aided, uplink relaying scheme using backscatter devices, and formulated a throughput maximization problem. Backscatter-enabled relays with energy harvesting capabilities were considered in \cite{SWIPTrelay, relayTA}, where \cite{SWIPTrelay} examined the performance of a relay capable of performing both simultaneous wireless information and power transfer (SWIPT) and backscatter communication from the source signal, where the relay utilized the conventional decode-and-forward (DF) scheme; and \cite{relayTA} derived the optimal time-switching schemes at an energy-harvesting relay capable of both backscatter and active transmissions. Work in \cite{threshold} derived detection thresholds and the bit error rate (BER) for a two-way backscatter communication system facilitated by a central relay, with the assumption that the receiver has knowledge of the channel state information (CSI). More recently, a system with an active source and a hybrid relay capable of both active and backscatter communication was studied in \cite{ambrelay}. The relay was powered by a field of energy sources while subjected to a separate group of interferers, and the coverage probability was derived for the ambient backscatter mode. The hybrid relay use case was explored further in \cite{Gong19} over a throughput maximization problem.

With the exception of \cite{threshold}, all of the above-mentioned works on backscatter relaying have considered either the equivalent of the conventional AF scheme, or hybrid devices with both active and backscatter transceivers. To the best of our knowledge, a comprehensive study of the backscatter-equivalent version of DF relaying is still lacking. While \cite{Mun19} suggested that it is not necessary to decode information prior to backscattering, we argue that for backscatter devices to be integrated into future wireless networks in their own right, the decoding function is necessary --- in fact critical, if upper-layer functionalities are required. Backscatter DF relaying would also facilitate local storage and processing of signals for future transmissions, in addition to future-proofing backscatter for applications where memory is required, such as the offloading of computation to more capable nodes and lightweight blockchain applications. Moreover, despite related works having extensively studied metrics such as throughput and coverage, the current literature lacks a fundamental BER characterization of the AF and DF backscatter relaying schemes individually, in addition to a comparison between the two schemes. While many comparisons of AF and DF relaying exist for active transceivers, it is not yet known whether the known results also generalize to passive relays. The theoretical insights obtained from such an analysis can be valuable for choosing an appropriate, application-dependent relaying scheme for future backscatter networks.

\subsection{Our Work and Contributions}

In this paper, we consider a two-hop relaying system where the relay's transmission is backscattering by nature. We demonstrate the feasibility of the backscatter relay under a practical system of study. The system model we consider is particularly applicable to blind spot scenarios where the presence of obstacles prevents direct communication between the source and destination. Such situations may occur in industrial and urban environments where it is desirable to provide coverage to the blind spot in a low-maintenance manner. We consider the presence of ambient interference at both the relay and destination, and characterize the BER performance under both DF and AF schemes to provide insights on the set of conditions where each scheme outperforms the other.

The main contributions of this paper are as follows:
\begin{itemize}
\item We introduce a new DF scheme specifically for backscatter relaying, where the relay's communication is assisted by the source node. A corresponding transmit power allocation problem is formulated for the source node, which is subject to a power budget constraint.
\item We derive the test statistics for energy-based detection of on-off keying (OOK) modulated signals at both the backscatter relay and the destination, in addition to the optimal and low-complexity detectors. The performance of the detector based on a Gaussian approximation of the detection statistic is shown to have good agreement with the optimal detector for BER up to $10^{-3}$.
\item We derive the analytical BER expressions for both the DF and AF relaying schemes, and examine the choice of reflection coefficients on the performance of each scheme. We find that due to the increased complexity brought about by decoding, the DF scheme exhibits worse performance compared to the AF scheme in the presence of  imperfections at the device level; however, both schemes perform similarly under ideal conditions.
\item Extensive numerical results on the outage probability performance for the backscatter relaying system under Rician fading are presented to demonstrate the feasibility of the proposed schemes, and to provide design insights on scenarios where each scheme may be more suitable.
\end{itemize}

\subsection{Paper Organization and Notations}

The rest of this paper is organized as follows. Section II introduces the system model. Section III presents the signal model for the DF and AF relaying schemes. Section IV derives the statistics required for detection at both relay and destination, and presents the detection thresholds. Section V presents the BER expressions and the source power allocation problem under the DF scheme. Numerical results are presented in Section VI and Section VII concludes the paper.

\textit{Notations}: We denote the expectation and variance operators by $\mathbb{E}\left\{\cdot\right\}$ and $\textrm{Var}\left\{\cdot\right\}$, respectively. $\mathbb{P}\left(\cdot\right)$ denotes the probability of an event. For complex-valued quantities, $\left|\cdot\right|$ denotes the magnitude, ${}^{*}$ denotes the complex conjugate, and $\textrm{Re}\left\{\cdot\right\}$ denotes the real part. $\mathcal{N}(\mu, \sigma^{2})$ and $\mathcal{CN}(\mu, \sigma^{2})$ represent Gaussian and complex Gaussian distributions, respectively, with mean $\mu$ and variance $\sigma^{2}$. $\Gamma(k, \theta)$ represents a gamma distribution with shape factor $k$ and scale factor $\theta$; while NC-$\chi^{2} (k; \lambda)$ represents a noncentral chi-squared distribution with $k$ degrees of freedom and noncentrality parameter $\lambda$.


\section{System Model}

We consider a system with three nodes: source, backscatter relay and destination, denoted by $S$, $R$ and $D$ respectively in subscripts hereafter. The source is an active radio with its own power supply; the relay transmits using backscatter modulation only; and the destination recovers the information from the relay. We  consider the presence of ambient interference signals originating from outside the system, which are received at both relay and destination, and denoted by $z_{R}[n]$ and $z_{D}[n]$, respectively. In addition, the noise terms at the relay and destination are denoted by $w_{R}[n]$ and $w_{D}[n]$, respectively. Note that the interference and noise are modeled separately, as the noise powers are similar at the relay and destination, but the interference powers can be significantly different between the two nodes. The system setup is shown in Fig. 1.

\begin{figure}
\centerline{\includegraphics[width=3.5in]{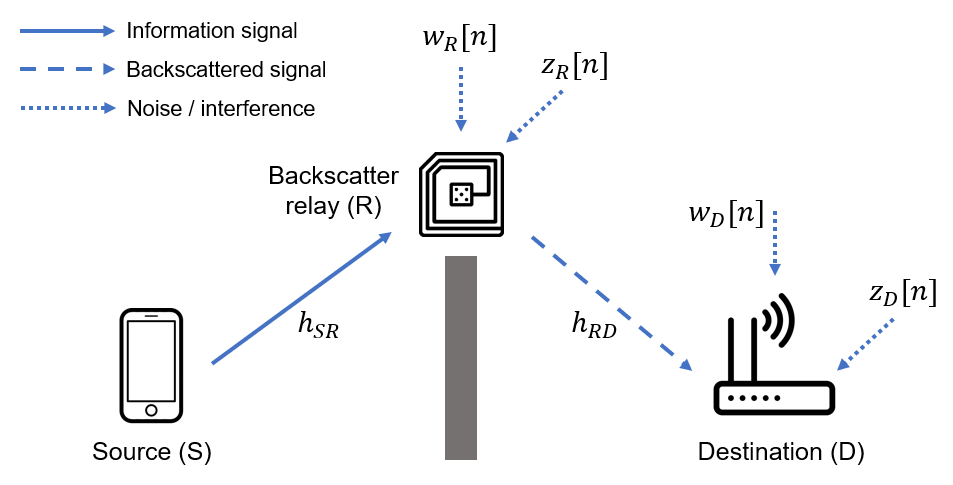}}
\caption{System model for backscatter relay-aided communication.\label{fig_1}}
\end{figure}

The source wishes to communicate to the destination assisted by the backscatter relay. We consider a blind spot scenario, where the presence of obstacles severely blocks the direct communication between the source and destination. Although such a blind spot assumption is not the most common scenario, it is particularly challenging to address, and hence requires special attention. In fact, this scenario has been adopted in existing backscatter communication literature \cite{ambrelay}, and more commonly in conventional relay networks and industrial IoT settings \cite{multihopaf, 2hinterferer}. The authors in \cite{Xu18} point out the potential of using backscatter devices to perform invasive monitoring tasks. One example is structural monitoring, where unfavorable propagation conditions may be experienced. We present a numerical case study in Section VI based on this scenario, to demonstrate the practicality of the assumption.

In the DF scheme, the relay decodes the signal from the source, and re-transmits it via backscattering. Each transmission occurs over two timeslots. In the first timeslot, the source transmits its data to the relay, which detects the received symbols. In the second timeslot, the source transmits a CW signal to support the backscatter transmission of the relay's received symbols to the destination. Note that the source's transmission of the CW signal in the second timeslot is similar to that in bistatic scatter in \cite{bistatic}, but is a unique feature when considered jointly with the backscatter relay system.

From a complexity perspective, the addition of decoding functionality to a backscatter device has been demonstrated in conventional RFID systems, where command signals from the reader are decoded at the tag. One method is to use variations of an envelope averaging circuit \cite{liu2013ambient}, which comprises of diodes, resistances, capacitances and a comparator. The overall power consumption of such devices is in the order of tens of $\mu$W \cite{Dob08}. More recently, works such as \cite{Sam08} and \cite{Nik12} have proposed tag prototypes where ultra-low-power microcontrollers with analog-to-digital converters are integrated with the antenna circuit, giving rise to the possibility of performing sampling and digital operations on received signals. The total power consumption of these prototypes has been demonstrated to be on the order of $1$ mW \cite{Sam08}, which can be readily sustained using a small battery. In comparison, the power consumption of active relays, as presented by \cite{Li07} and \cite{Mah15}, is up to several hundred mW. Therefore, the DF backscatter operation is justified by its ultra-low-power implementation of decoding circuits. In this paper, we assume the relay to be semipassive and taking on the tag architecture in \cite{Sam08, Nik12}.

In the AF scheme, the incoming signal from the source is directly backscattered by the relay without delay. Hence, the destination receives the backscattered signal within the same timeslot as the transmission from the source.

While AF backscatter relaying was considered in \cite{BFRelay}, the BER performance was not derived therein. Hereafter, we consider the AF scheme as the baseline, and compare it with the BER performance of the DF scheme to characterize the differences attributed to the decoding operation. Note that the AF scheme technically does not amplify the signal, as the relay is not able to increase the signal power through reflection. We refer to the scheme as AF for convenience.

In this paper, we consider BER purely based on the bits transmitted and received. The performance gains brought about by error correction codes is outside the scope of this work. We assume OOK modulation, where the source performs active transmission and the relay performs backscatter modulation. The use of OOK is consistent with backscatter literature, and is necessary to ensure maximum probability of reception under non-line-of-sight situations.


\section{Signal Model and Relaying Protocols}
We consider a baseband discrete-time signal model where the time index of the signal samples is denoted by $n$. The sampling rate for signal reception is set according to the rate of change (or the equivalent symbol rate) of the ambient interference signal. On the other hand, each data symbol transmitted by the source and backscattered by the relay spans $N$ samples. In other words, the data symbol rate is much lower than the sampling rate at the receiver end. This is in agreement with the signal model used in conventional ambient backscatter works such as \cite{Amb16, noncoherent, exactBER}, where the data symbol rate is intentionally reduced compared to the sampling rate.

Once the sampling rate and symbol rate are chosen, they are fixed for all transmissions. Although the spectral efficiency of the transmissions is reduced compared to conventional communications, the resulting low data rate is appropriate for most applications of backscatter communications, including our scenario and those in e.g. \cite{Amb16, noncoherent, exactBER}. Here, often the key objective is to achieve targeted reliability over the desired communication range, without requiring excessive power consumption (e.g., to report a small amount of sensed data from an IoT device). Therefore, low-rate transmission is often adopted in order to achieve the required reliability.

The channel model accounts for both small-scale fading and path loss. We let the links between the relaying devices take on a line-of-sight component, and assume Rician fading channels with quasi-static block fading. The channel coefficient between nodes $a$ and $b$ is given by $h_{ab} = \sqrt{l_{ab}} h'_{ab}$, where $h'_{ab} \sim \mathcal{CN}( \sqrt{\frac{K}{K+1}}, \frac{1}{K+1} )$, and $K$ is the Rician $K$-factor. The path loss $l_{ab}$ can be modeled by $\frac{G_{t} G_{r} (c/f_{c})^{2}}{d^{\gamma} (4 \pi)^{2}}$, where $G_{t}$ and $G_{r}$ are the antenna gains at the transmitter and receiver, respectively; $c = 3 \times 10^{8} m/s$; $f_{c}$ being the carrier frequency; $d$ denoting distance between nodes; and $\gamma$ being the path loss exponent.

\subsection{Backscatter Operation}

In the DF scheme, the backscatter relay performs modulation by switching between two load impedances connected to the antenna, each corresponding to a reflection coefficient that determines the amount of reflected power. The reflection coefficient is denoted by $\Gamma[n]$, and is constant over each source symbol period of $N$ samples. The use of two impedances under binary modulation results in two reflection coefficients $\Gamma_{0}$ and $\Gamma_{1}$ which are the two possible values of $\Gamma[n]$. The baseband signal at the relay is given by
\begin{equation}
B[n] = A - \Gamma[n],	\label{structuralMode}
\end{equation}
where $A$ is a term related to the antenna structural mode \cite{bistatic}.\footnote{The structural mode is a constant depending on the geometry and construction of the antenna, and not the antenna's operating environment.} Here, we assume the relay has a general (non-minimum-scattering) antenna, where $A$ is complex valued \cite{improving}; further, $|A| \leq 1$. We let $\Gamma_{0}$ and $\Gamma_{1}$ take on general complex values satisfying $|\Gamma_{0}|, |\Gamma_{1}| \leq 1$. Moreover, we denote $B_{0}$ and $B_{1}$ as the two values of $B[n]$ corresponding to reflection coefficients $\Gamma_{0}$ and $\Gamma_{1}$, respectively. We assume that both relay and destination have knowledge of the transmitted bit corresponding to each energy level in its received signal, to account for the possibility that bit $1$ corresponds to the lower energy level, and vice versa. This is achieved through the use of pilot sequences, which are not explicitly considered here.

\subsection{Interference Modeling}

We denote the interference received by node $i \in \{R,D\}$ in timeslot $j \in \{1,2\}$ as $\sqrt{P_{I,i}} z_{i,j}[n]$, where $P_{I,i}$ is the received interference power, and $z_{i,j}[n]$, for all $n$, are independent and identically distributed (i.i.d.) interference samples following $\mathcal{CN}(0,1)$. Realistically, the signals transmitted by each interferer would use well-defined modulation schemes. As such, the exact distribution of the interference signal would not be Gaussian. However, it is often impractical to obtain the statistical description of every ambient signal. Hence, the use of the Gaussian assumption is a reasonable simplification to approximate the overall distribution of the aggregate interference. Note that the Gaussian assumption was also used in \cite{noncoherent} to model the distribution of the ambient signal.

Due to the nature of backscatter relaying, the interference received at both relay and destination jointly affects the detection at the destination. We consider a  signal model where $z_{R,j}[n]$ and $z_{D,j}[n]$ are uncorrelated (i.e. the signals that make up $z_{R,j}[n]$ and $z_{D,j}[n]$ are unique to each receiver), in order to capture the variable nature of ambient interference signals received at different locations.

\subsection{The DF Scheme}

As mentioned, two timeslots are required for each transmission under the DF scheme. The signal received at the relay in the first timeslot is given by
\begin{equation}
y_{R}[n] = \sqrt{P_{S,1}} h_{SR} x[n] + \sqrt{P_{I,R}} z_{R,1}[n] + w_{R}[n],	\label{received_DFR}
\end{equation}
where $P_{S,1}$ is the source transmit power in the first timeslot, $x[n] \in \{0, 1\}$ is the OOK-modulated source data symbol, and $w_{R}[n]$ is noise at the relay, following $\mathcal{CN}(0, P_{w,R})$.\footnote{In this paper, we assume that the low-pass filtering bandwidth at each receiver is appropriately chosen according to the sampling rate, and is fixed once chosen, such that the noise samples are uncorrelated over all transmissions. This is in agreement with the noise process modeling in \cite{bistatic, Amb16, noncoherent, exactBER}.} We let the source transmit a CW signal in the second timeslot with power $P_{S,2}$ to boost the backscatter transmission at the relay. Following demodulation in the first timeslot, the relay backscatters the incoming signal, consisting of the CW signal from the source and the interference. Hence, the signal backscattered from the relay in the second timeslot is
\begin{equation}
x_{R}[n] = \eta \left( \sqrt{P_{S,2}} h_{SR} + \sqrt{P_{I,R}} z_{R,2}[n] \right) B[n].	\label{sent_DFR}
\end{equation}
Here, $\eta \in (0, 1]$ is the backscatter switching loss coefficient and is modeled as a constant. In turn, the destination receives
\begin{align}
y_{D}[n] &= h_{RD} x_{R}[n] + \sqrt{P_{I,D}} z_{D,2}[n] + w_{D}[n] \nonumber \\
		&= \eta h_{RD} \left( \sqrt{P_{S,2}} h_{SR} + \sqrt{P_{I,R}} z_{R,2}[n] \right) B[n] + \sqrt{P_{I,D}} z_{D,2}[n] + w_{D}[n],	\label{received_DFD}
\end{align}
where $w_{D}[n] \sim \mathcal{CN}(0, P_{w,D})$ is the noise at the destination.

\subsection{The AF Scheme}

In the full-duplex AF scheme, the end-to-end transmission is completed in one timeslot. The signal received at the relay is similar to (3) and is given by
\begin{equation}
y_{R}[n] = \sqrt{P_{S}} h_{SR} x[n] + \sqrt{P_{I,R}} z_{R,1}[n].	\label{received_AFR}
\end{equation}
Here, the source transmit power is denoted by $P_{S}$. Note that (\ref{received_AFR}) does not include a noise term, similar to the signal model presented in \cite{BFRelay}. This is because the received signal is directly backscattered by the relay and does not undergo any processing. The baseband signal at the relay has the same form as (\ref{structuralMode}); however, for the AF operation, the backscatter reflection coefficient $\Gamma[n]$ is set to the one which results in the larger magnitude of $B[n]$. The larger reflection coefficient is denoted by $\Gamma$.\footnote{More specifically, $\Gamma = \Gamma_{0}$ if $|A - \Gamma_{0}| > |A - \Gamma_{1}|$, and $\Gamma = \Gamma_{1}$ otherwise.} Since $\Gamma[n]$ is a constant, we drop the indexing of $B[n]$ and write the baseband signal as $B \triangleq A - \Gamma$ for the entire transmission. The signal received at the destination is
\begin{align}
y_{D}[n] &= \eta h_{RD} B y_{R}[n] + \sqrt{P_{I,D}} z_{D,1}[n] + w_{D}[n] \nonumber \\
         &= \eta \sqrt{P_{S}} h_{SR} h_{RD} B  x[n] + \eta \sqrt{P_{I,R}} h_{RD} B  z_{R,1}[n] + \sqrt{P_{I,D}} z_{D,1}[n] + w_{D}[n],	\label{received_AFD}
\end{align}
where in (\ref{received_AFD}), the first two terms represent the signal backscattered by the relay, and the third term is the interference received at the destination.

\subsection{Performance Metrics}

We use the BER as the main metric when considering performance characterization. In addition, we consider the outage probability as a long-term measure, where an outage occurs whenever the BER within any channel coherence period exceeds some threshold. The results on the outage probability are presented in Section VI. It should be noted that the performance comparison of AF and DF schemes in terms of BER or outage probability does not take into account the difference in transmission time or throughput. However, for applications requiring only low-rate transmission of fixed-size or short packets, reliability is the main concern, whereas throughput is a less relevant metric.


\section{Energy-Based Detection}

We consider energy-based detection at both the relay and the destination. An averaging circuit collects the received signal samples over the length of one data symbol ($N$ ambient samples/symbols) and obtains the average received signal power over all samples. This quantity is referred to as the test statistic throughout this section. The averaging mechanism is required to tackle the noise and interference at each receiver. The received bit is determined by comparing the test statistic with a detection threshold, which depends on the distributions of the test statistic corresponding to each bit. In this section, we derive both exact and approximate probability density functions (pdfs) of the test statistic conditioned on the bit sent, in addition to their respective detection thresholds.

\subsection{Detection Statistics at the Relay for the DF Scheme}

The average power over one symbol period of the received signal at the relay, $y_{R}[n]$, namely the test statistic $\psi^{DF,R}$, is given by
\begin{equation}
\psi^{DF,R} = \frac{1}{N} \sum_{n=0}^{N-1} \left|\sqrt{P_{S,1}} h_{SR} x[n] + \sqrt{P_{I,R}} z_{R,1}[n] + w_{R}[n]\right|^{2}.	\label{TS_DFR}
\end{equation}
Note that $\psi^{DF,R}$ takes on two values: $\psi^{DF,R}_{0}$ and $\psi^{DF,R}_{1}$, corresponding to the cases where $x[n] = 0$ and $x[n] = 1$, respectively. Then we have 
\begin{subequations}
\begin{align}
\psi^{DF,R}_{0} &= \sum_{n=0}^{N-1} \frac{P_{I,R} |z_{R,1}[n]|^{2} + |w_{R}[n]|^{2}}{N} + \sum_{n=0}^{N-1} \frac{2\textrm{Re}\left\{\sqrt{P_{I,R}}  z_{R,1}[n] w_{R}[n]^{*}\right\}}{N}, \label{DFR_a} \\
\psi^{DF,R}_{1} &= \sum_{n=0}^{N-1} \frac{P_{I,R} |z_{R,1}[n]|^{2} + P_{S,1} |h_{SR}|^{2} + |w_{R}[n]|^{2}}{N} + \sum_{n=0}^{N-1} \frac{2\textrm{Re}\left\{\sqrt{P_{I,R}} z_{R,1}[n] w_{R}[n]^{*}\right\}}{N} \nonumber \\
        &\qquad + \sum_{n=0}^{N-1} \frac{2\textrm{Re}\left\{\sqrt{P_{S,1} P_{I,R}}  h_{SR} z_{R,1}[n]^{*}\right\}}{N} + \sum_{n=0}^{N-1} \frac{2\textrm{Re}\left\{\sqrt{P_{S,1}} h_{SR} w_{R}[n]^{*}\right\}}{N}. \label{DFR_b}
\end{align}
\end{subequations}
Expanding (\ref{TS_DFR}) in terms of its real and imaginary components and evaluating the distribution of each component yields the following proposition.

\begin{proposition}
The two values of the test statistic $\psi^{DF,R}$ can be modeled as random variables 
\begin{equation}
\psi^{DF,R}_{0} \sim \Gamma\left(N, \frac{\sigma_{DF}^{2}}{N}\right), \hspace{5mm} \psi^{DF,R}_{1} \sim \text{NC-}\chi^{2}\left(k = 2N; \lambda = \frac{2 N P_{S,1} |h_{SR}|^{2}}{\sigma_{DF}^{2}}\right), \nonumber
\end{equation}
and their exact pdfs, denoted by $f^{DF,R}_{\psi_{0}}(x)$ and $f^{DF,R}_{\psi_{1}}(x)$, are given by
\begin{align}
f^{DF,R}_{\psi_{0}}(x) &= \frac{1}{\Gamma(N)} \left( \frac{N}{\sigma_{DF}^{2}} \right)^{N} x^{N-1} \exp \left( -\frac{N x}{\sigma_{DF}^{2}} \right), \label{prop1a} \\
f^{DF,R}_{\psi_{1}}(x) &= \frac{N}{\sigma_{DF}^{2}} \exp \left( -\frac{N}{\sigma_{DF}^{2}} \left( x + P_{S,1} |h_{SR}|^{2} \right) \right) \left( \frac{x}{P_{S,1} |h_{SR}|^{2}} \right)^{\frac{N-1}{2}} \nonumber \\ & \qquad \times I_{N-1} \left( \frac{2 N}{\sigma_{DF}^{2}} \sqrt{P_{S,1} |h_{SR}|^{2} x} \right), \label{prop1b}
\end{align}
where $\Gamma(\cdot)$ denotes the gamma function, $I_{\nu}(\cdot)$ denotes the modified Bessel function of the first kind with order $\nu$, and $\sigma_{DF}^{2} = P_{I,R} + P_{w,R}$ is common to both (\ref{prop1a})-(\ref{prop1b}).
\end{proposition}

\noindent \textit{Proof.} The main steps for the derivation of (\ref{prop1a}) is presented in Appendix A. The derivation of (\ref{prop1b}) is similar to the steps given in \cite[Appendix A]{exactBER}; however, for completeness, the main steps are also presented in Appendix A.

When $N$ is large, we can invoke the central limit theorem (CLT) on (\ref{DFR_a})-(\ref{DFR_b}) to obtain a Gaussian approximation of the test statistic $\psi^{DF,R}$, given in the following proposition.
\begin{proposition} Using the CLT, as $N \rightarrow \infty$,
\begin{equation}
\psi_{0}^{DF,R} \sim \mathcal{N}(\mu_{0}, \hat{\sigma}_{0}^{2}), \hspace{10mm} \psi_{1}^{DF,R} \sim \mathcal{N}(\mu_{1}, \hat{\sigma}_{1}^{2}),	\label{prop2a}
\end{equation}
where the mean values are
\begin{subequations}
\begin{align}
\mu_{0} &= \mathbb{E}\{\psi_{0}^{DF,R}\} = \sigma_{DF}^{2}, \\
\mu_{1} &= \mathbb{E}\{\psi_{1}^{DF,R}\} = P_{S,1} |h_{SR}|^{2} + \sigma_{DF}^{2},	\label{prop2b}
\end{align}
\end{subequations}
and the variances are
\begin{equation}
\hat{\sigma}_{0}^{2} = \frac{\left(\sigma_{DF}^{2}\right)^{2}}{N},  \hspace{10mm} \hat{\sigma}_{1}^{2} = \frac{2 P_{S,1} |h_{SR}|^{2} \sigma_{DF}^{2}}{N} + \hat{\sigma}_{0}^{2}.	\label{prop2c}
\end{equation}
\end{proposition}

\noindent \textit{Proof.} Equation (\ref{prop2c}) can be readily derived by calculating the variance for each individual term in (\ref{DFR_a})-(\ref{DFR_b}), and then summing the variances over all terms.

\subsection{Detection Statistics at the Destination for the DF Scheme}

Given the backscatter modulation at the relay, the test statistic at the destination, denoted by $\psi^{DF,D}$, depends on the relay baseband signal $B[n]$. The average power of the received signal $y_{D}[n]$ at the destination over one symbol is given by 
\begin{align}
\psi^{DF,D} &= \frac{1}{N} \sum_{n=0}^{N-1} \left|\eta h_{RD} \left( \sqrt{P_{S,2}} h_{SR} + \sqrt{P_{I,R}} z_{R,2}[n] \right) B[n] + \sqrt{P_{I,D}} z_{D,2}[n] + w_{D}[n]\right|^{2}.	\label{TS_DFD}
\end{align}
Note that $\psi^{DF,D}$ takes on two values, $\psi^{DF,D}_{0}$ and $\psi^{DF,D}_{1}$, corresponding to the cases where the backscattered bit by the relay is 0 and 1, respectively. Then we have
\begin{align}
\psi^{DF,D}_{i} &= \frac{1}{N} \sum_{n=0}^{N-1} \left|\alpha_{DF,i} + \beta_{DF,i} z_{R,2}[n] + \sqrt{P_{I,D}} z_{D,2}[n] + w_{D}[n] \right|^{2}, \label{DFD}
\end{align}
where we have defined $\alpha_{DF,i} = \eta \sqrt{P_{S,2}} h_{SR} h_{RD} B[n]$, and $\beta_{DF,i} = \eta \sqrt{P_{I,R}} h_{RD} B[n]$, for $x[n] = i \in \{0, 1\}$. Expanding (\ref{DFD}) into its real and imaginary components yields the following expressions for the pdfs of $\psi^{DF,D}_{0}$ and $\psi^{DF,D}_{1}$.

\begin{proposition}
The two values of the test statistic, $\psi^{DF,D}_{i}, \ i \in \{0, 1\}$, can be modeled as random variables 
\begin{equation}
\psi^{DF,D}_{i} \sim \text{NC-}\chi^{2} \left( k = 2N; \lambda = \frac{2 N |\alpha_{DF,i}|^{2}}{\sigma_{i}^{2}} \right), \nonumber
\end{equation}
and their exact pdfs are given by
\begin{equation}
f^{DF,D}_{\psi_{i}}(x) = \frac{N}{\sigma_{i}^{2}} \exp \left( -\frac{N}{\sigma_{i}^{2}} \left( x + |\alpha_{DF,i}|^{2} \right) \right) \left( \frac{x}{|\alpha_{DF,i}|^{2}} \right)^{\frac{N-1}{2}} I_{N-1} \left( \frac{2 N |\alpha_{DF,i}|}{\sigma_{i}^{2}} \sqrt{x} \right),	\label{prop3}
\end{equation}
where $\sigma_{i}^{2} = |\beta_{DF,i}|^{2} + P_{I,D} + P_{w,D}$.
\end{proposition}

\noindent \textit{Proof.} The derivation is analogous to the steps given for $f^{DF,R}_{\psi_{1}}(x)$ in Appendix A.

Again, we can invoke the CLT on (\ref{DFD}) to obtain the Gaussian approximation of the test statistic $\psi^{DF,D}$, given in the following proposition.

\begin{proposition}
Using the CLT, as $N \rightarrow \infty$,
\begin{equation}
\psi^{DF,D}_{0} \sim \mathcal{N}(\mu_{0}, \hat{\sigma}_{0}^{2}), \hspace{10mm} \psi^{DF,D}_{1} \sim \mathcal{N}(\mu_{1}, \hat{\sigma}_{1}^{2}),	\label{prop4a}
\end{equation}
where, for $i \in \{0, 1\}$, the mean values are
\begin{equation}
\mu_{i} = |\alpha_{DF,i}|^{2} + \sigma_{i}^{2},	\label{prop4b}
\end{equation}
and the variances are
\begin{equation}
\hat{\sigma}_{i}^{2} = \frac{\left( \sigma_{i}^{2} \right)^{2} + 2 |\alpha_{DF,i}|^{2} \sigma_{i}^{2}}{N}.	\label{prop4c}
\end{equation}
\end{proposition}

\subsection{Detection Statistics at the Destination for the AF Scheme}

The average power of the signal received by the destination over one source symbol under the AF scheme, namely the test statistic $\psi^{AF}$, is given by
\begin{equation}
\psi^{AF} = \frac{1}{N} \sum_{n=0}^{N-1} \left|\eta h_{RD} B \left(\sqrt{P_{S}} h_{SR} x[n] + \sqrt{P_{I,R}} z_{R,1}[n]\right) + \sqrt{P_{I,D}} z_{D,1}[n] + w_{D}[n]\right|^{2}.	\label{TS_AFD}
\end{equation}
Similar to the test statistics for the DF scheme, the test statistic $\psi^{AF}$ at the destination takes on two values $\psi^{AF}_{0}$ and $\psi^{AF}_{1}$, corresponding to $x[n] = 0$ and $x[n] = 1$, respectively, given as follows:
\begin{subequations}
\begin{align}
\psi^{AF}_{0} &= \frac{1}{N} \sum_{n=0}^{N-1} \left|\beta_{AF} z_{R,1}[n] + \sqrt{P_{I,D}} z_{D,1}[n] + w_{D}[n] \right|^{2}, \label{AFD_a} \\
\psi^{AF}_{1} &= \frac{1}{N} \sum_{n=0}^{N-1} \left|\alpha_{AF} + \beta_{AF} z_{R,1}[n] + \sqrt{P_{I,D}} z_{D,1}[n] + w_{D}[n] \right|^{2}, \label{AFD_b}
\end{align}
\end{subequations}
where $\alpha_{AF} = \eta \sqrt{P_{S}} h_{SR} h_{RD} B$, and $\beta_{AF} = \eta \sqrt{P_{I,R}} h_{RD} B$. The exact distribution of $\psi^{AF}$ and the Gaussian approximations are given in the following two propositions.

\begin{proposition}
The two values of the test statistic $\psi^{AF}$ can be modeled as random variables 
\begin{equation}
\psi^{AF}_{0} \sim \Gamma\left(N, \frac{\sigma_{AF}^{2}}{N}\right), \hspace{5mm} \psi^{AF}_{1} \sim \text{NC-}\chi^{2}\left(k = 2N; \lambda = \frac{2 N |\alpha_{AF}|^{2}}{\sigma_{AF}^{2}}\right), \nonumber
\end{equation}
and their exact pdfs are given by
\begin{align}
f_{\psi_{0}}^{AF}(x) &= \frac{1}{\Gamma(N)}  \left( \frac{N}{\sigma_{AF}^{2}} \right)^{N} x^{N-1} \exp \left( -\frac{N x}{\sigma_{AF}^{2}} \right), \label{prop5a} \\
f_{\psi_{1}}^{AF}(x) &= \frac{N}{\sigma_{AF}^{2}} \exp \left( -\frac{N}{\sigma_{AF}^{2}} \left( x + |\alpha_{AF}|^{2} \right) \right) \left( \frac{x}{|\alpha_{AF}|^{2}} \right)^{\frac{N-1}{2}} \ I_{N-1} \left( \frac{2 N |\alpha_{AF}|}{\sigma_{AF}^{2}} \sqrt{x} \right),	\label{prop5b}
\end{align}
where $\sigma_{AF}^{2} = |\beta_{AF}|^{2} + P_{I,D} + P_{w,D}$ is common to both pdfs.
\end{proposition}

\begin{proposition}
Using the CLT, as $N \rightarrow \infty$, 
\begin{equation}
\psi^{AF}_{0} \sim \mathcal{N}(\mu_{0}, \hat{\sigma}_{0}^{2}), \hspace{10mm} \psi^{AF}_{1} \sim \mathcal{N}(\mu_{1}, \hat{\sigma}_{1}^{2}),	\label{prop6a}
\end{equation}
where the mean values are
\begin{equation}
\mu_{0} = \sigma_{AF}^{2}, \hspace{10mm} \mu_{1} = |\alpha_{AF}|^{2} + \sigma_{AF}^{2},	\label{prop6b}
\end{equation}
and the variances are
\begin{equation}
\hat{\sigma}_{0}^{2} = \frac{\left(\sigma_{AF}^{2}\right)^{2}}{N}, \hspace{10mm} \hat{\sigma}_{1}^{2} = \frac{2 |\alpha_{AF}|^{2} \sigma_{AF}^{2}}{N} + \hat{\sigma}_{0}^{2}.	\label{prop6c}
\end{equation}
\end{proposition}

It should be noted that slight abuses of notation have been used in Section IV-A to IV-C to denote the distribution parameters of the test statistics in different cases. Specifically, $\mu_{0}$ and $\mu_{1}$ denote the mean values for both the exact characterization and Gaussian approximation of the test statistic distributions. Moreover, $\sigma_{DF}$, $\sigma_{AF}$, $\sigma_{0}$ and $\sigma_{1}$ are parameters denoting certain second-order statistics for the exact test statistic distributions; whereas $\hat{\sigma}_{0}$ and $\hat{\sigma}_{1}$ denote the variances of the Gaussian approximations. These parameters have different expressions under both DF and AF schemes, as well as at the relay and destination.

\subsection{Detection Threshold}

The detected symbol at the relay or destination, denoted by $\hat{x}[n]$, is determined according to the following detection rule:
\begin{equation}
\hat{x}[n] = 
	\begin{cases}
		1, & \psi > T, \\
		0, & \psi < T,
	\end{cases}
\end{equation}
with $\psi$ being a test statistic from Section IV-A to IV-C depending on the receiver and relaying scheme, and $T$ being a detection threshold. The optimal detection thresholds for the test statistics of the DF scheme at the relay, DF scheme at the destination and AF scheme at the destination are obtained by equating the two pdf expressions $f_{\psi_{0}}(x)$ and $f_{\psi_{1}}(x)$ in Propositions 1, 3 and 5, respectively, in the same way as for binary modulation schemes using maximum likelihood (ML) detection, and in related backscatter works \cite{Amb16, exactBER}. The thresholds are summarized in the following result.

\begin{theorem}
The optimal detection thresholds for the DF scheme at the relay ($T_{DF,R}^{*}$), DF scheme at the destination ($T_{DF,D}^{*}$), and the AF scheme at the destination ($T_{AF}^{*}$), are the solutions to the following equations, respectively:
\begin{multline}
\frac{\pi}{\Gamma(n)} \left( \frac{N \sqrt{P_{S,1} |h_{SR}|^{2} T_{DF,R}^{*}}}{\sigma_{DF}^{2}} \right)^{N-1} \exp \left( \frac{N P_{S,1} |h_{SR}|^{2}}{\sigma_{DF}^{2}} \right) \\ = \int_{0}^{\pi} \exp \left( \frac{2N}{\sigma_{DF}^{2}} \sqrt{P_{S,1} T_{DF,R}^{*}} \cos(\theta) \right) \cos(N - 1)\theta \ \textrm{d}\theta,	\label{thm1a}
\end{multline}
\begin{multline}
\frac{\sigma_{1}^{2}}{\sigma_{0}^{2}} \left( \frac{|B_{1}|}{|B_{0}|} \right)^{N-1} \exp \left( \left( \frac{N}{\sigma_{1}^{2}} - \frac{N}{\sigma_{0}^{2}} \right) T_{DF,D}^{*} + \left( \frac{N |\alpha_{DF,1}|^{2}}{\sigma_{1}^{2}} - \frac{N |\alpha_{DF,0}|^{2}}{\sigma_{0}^{2}} \right)  \right) \\ \times \int_{0}^{\pi} \exp \left( \frac{2N |\alpha_{DF,0}|}{\sigma_{0}^{2}} \sqrt{T_{DF,D}^{*}} \cos(\theta) \right) \cos(N - 1)\theta \ \textrm{d}\theta \\ = \int_{0}^{\pi} \exp \left( \frac{2N |\alpha_{DF,1}|}{\sigma_{1}^{2}} \sqrt{T_{DF,D}^{*}} \cos(\theta) \right) \cos(N - 1)\theta \ \textrm{d}\theta,	\label{thm1b}
\end{multline}
\begin{multline}
\frac{\pi}{\Gamma(n)} \left( \frac{N |\alpha_{AF}| \sqrt{T_{AF}^{*}}}{\sigma_{AF}^{2}} \right)^{N-1} \exp \left( \frac{N |\alpha_{AF}|^{2}}{\sigma_{AF}^{2}} \right) \\
= \int_{0}^{\pi} \exp \left( \frac{2N |\alpha_{AF}|}{\sigma_{AF}^{2}} \sqrt{T_{AF}^{*}} \cos(\theta) \right) \cos(N - 1)\theta \ \textrm{d}\theta,	\label{thm1c}
\end{multline}
with $\sigma_{DF}^{2}$, $\sigma_{AF}^{2}$, $\sigma_{0}^{2}$, $\sigma_{1}^{2}$, $|\alpha_{DF,0}|$, $|\alpha_{DF,1}|$ and $|\alpha_{AF}|$ given in Section IV-I to IV-C.
\end{theorem}

\noindent \textit{Proof.} See Appendix B. Note that the derivation for (\ref{thm1b}) is similar to that presented in \cite[Appendix B]{exactBER}, and is provided in Appendix B for completeness.

The detection thresholds for the Gaussian approximation are obtained by solving the equation of the pdfs of the two random variables in (\ref{prop2a}), (\ref{prop4a}) and (\ref{prop6a}) similar to the procedure in \cite{threshold}. We present the complete version of the result as follows.

\begin{theorem}
For either source-to-relay or relay-to-destination links under the DF scheme and the source-to-destination link under the AF scheme, the Gaussian-approximated detection threshold $T_{G}$ takes on two possible values:
\begin{equation}
T_{G} = \frac{(\hat{\sigma}_{0}^{2} \mu_{1} - \hat{\sigma}_{1}^{2} \mu_{0})}{\hat{\sigma}_{0}^{2} - \hat{\sigma}_{1}^{2}} \pm \sqrt{\frac{\hat{\sigma}_{0}^{2} \hat{\sigma}_{1}^{2} \left( \left(\mu_{0} - \mu_{1}\right)^{2} + 2 \left(\hat{\sigma}_{0}^{2} - \hat{\sigma}_{1}^{2}\right) \ln\left(\frac{\hat{\sigma}_{0}}{\hat{\sigma}_{1}}\right)\right)}{\left(\hat{\sigma}_{0}^{2} - \hat{\sigma}_{1}^{2}\right)^{2}}},	\label{thm2}
\end{equation}
where the solution with the positive sign is taken if $\mu_{1} > \mu_{0}$, and the solution with the negative sign is taken otherwise.
\end{theorem}

\noindent \textit{Proof.} See Appendix C.

Note that the optimal detection thresholds for both the exact distribution and Gaussian approximation require different levels of knowledge about the test statistic for each symbol. In the worst-case scenario, where no statistical knowledge is available, we propose a simple threshold derived by taking the average of the entire set of test statistic values corresponding to all received symbols, which mathematically equates to
\begin{equation}
T_{S} \triangleq \frac{\mu_{0} + \mu_{1}}{2}.	\label{simpleThreshold}
\end{equation}


\section{BER Performance and Source Power Optimization}

\subsection{BER Performance}

The BER expression for binary modulation over a single link is given by
\begin{align}
p_{b} \ & = \mathbb{P}(x[n] = 0) \ \mathbb{P}(\hat{x}[n] = 1 | x[n] = 0) \nonumber \\
		& + \mathbb{P}(x[n] = 1) \ \mathbb{P}(\hat{x}[n] = 0 | x[n] = 1).	\label{singleLinkBER}
\end{align}
Here, $\mathbb{P}(\hat{x}[n] = 1 | x[n] = 0)$ and $\mathbb{P}(\hat{x}[n] = 0 | x[n] = 1)$ denote the incorrect detection probabilities when bits $0$ and $1$ are sent, respectively. In our system, they are equivalent to the integrals of $f_{\psi_{0}}(x)$ and $f_{\psi_{1}}(x)$ over the values of $x$ on the opposite side of a detection threshold (i.e. $T_{E}$, $T_{G}$ or $T_{S}$). Under the DF scheme, the end-to-end BER, denoted by $p_{b}^{DF}$, is a function of the two individual BERs for the source-to-relay and relay-to-destination links, and was shown in \cite{highDF} to be
\begin{align}
p_{b}^{DF} &= p_{b}^{(1)} \left( 1 - p_{b}^{(2)} \right) + p_{b}^{(2)} \left( 1 - p_{b}^{(1)} \right) \nonumber \\
		&= p_{b}^{(1)} + p_{b}^{(2)} - 2 p_{b}^{(1)} p_{b}^{(2)},	\label{twoLinkBER}
\end{align}
where $p_{b}^{(1)}$ and $p_{b}^{(2)}$ are the source-to-relay and relay-to-destination BERs, respectively. The BER for the AF scheme, denoted by $p_{b}^{AF}$, can be calculated directly using (\ref{singleLinkBER}).

For each link, two expressions for the BER exist, for when $\mu_{0} < \mu_{1}$ and $\mu_{0} > \mu_{1}$. We denote the two cases using subscripts $a$ and $b$ in the following equations.

\subsubsection{DF scheme} The exact BER expressions for the source-to-relay link using the optimal detection threshold can be written as
\begin{subequations}
\begin{align}
p_{b}^{opt,1a} &= \frac{1}{2} \left[ 2 - \frac{1}{\Gamma(N)} \gamma\left( N, \frac{N T_{DF,R}^{*}}{\sigma^{2}} \right) - Q_{N} \left( \sqrt{\frac{2 N P_{S,1} |h_{SR}|^{2}}{\sigma^{2}}}, \sqrt{\frac{2 N T_{DF,R}^{*}}{\sigma^{2}}} \right) \right], \label{BER_DF1a} \\
p_{b}^{opt,1b} &= \frac{1}{2} \left[ \frac{1}{\Gamma(N)} \gamma\left( N, \frac{N T_{DF,R}^{*}}{\sigma^{2}} \right) + Q_{N} \left( \sqrt{\frac{2 N P_{S,1} |h_{SR}|^{2}}{\sigma^{2}}}, \sqrt{\frac{2 N T_{DF,R}^{*}}{\sigma^{2}}} \right) \right], \label{BER_DF1b}
\end{align}
\end{subequations}
where $\gamma(a, x) = \int_{0}^{x} t^{a-1} e^{-t} \ \textrm{d}t$ and $Q_{M}(a, b) = \int_{b}^{\infty} x \left( \frac{x}{a} \right)^{M-1} \exp( -\frac{x^{2} + a^{2}}{2} ) I_{M-1}(ax) \ \textrm{d}x$ denote the lower incomplete gamma function and the Marcum Q-function, respectively. Similarly, for the relay-to-destination link, we have
\begin{subequations}
\begin{align}
p_{b}^{opt,2a} &= \frac{1}{2} \left[ 1 + Q_{N} \left( \sqrt{\frac{2 N |\alpha_{DF,0}|^{2}}{\sigma_{0}^{2}}}, \sqrt{\frac{2 N T_{DF,D}^{*}}{\sigma_{0}^{2}}} \right) - Q_{N} \left( \sqrt{\frac{2 N |\alpha_{DF,1}|^{2}}{\sigma_{1}^{2}}}, \sqrt{\frac{2 N T_{DF,D}^{*}}{\sigma_{1}^{2}}} \right) \right], \label{BER_DF2a} \\
p_{b}^{opt,2b} &= \frac{1}{2} \left[ 1 + Q_{N} \left( \sqrt{\frac{2 N |\alpha_{DF,1}|^{2}}{\sigma_{1}^{2}}}, \sqrt{\frac{2 N T_{DF,D}^{*}}{\sigma_{1}^{2}}} \right) - Q_{N} \left( \sqrt{\frac{2 N |\alpha_{DF,0}|^{2}}{\sigma_{0}^{2}}}, \sqrt{\frac{2 N T_{DF,D}^{*}}{\sigma_{0}^{2}}} \right) \right]. \label{BER_DF2b}
\end{align}
\end{subequations}

\subsubsection{AF scheme} Since the relay does not process the source information signal, the source-to-relay and relay-to-destination links can effectively be considered as one link. The exact BER expression based on the optimal detection threshold is given by
\begin{subequations}
\begin{align}
p_{b}^{opt,AFa} &= \frac{1}{2} \left[ 2 - \frac{1}{\Gamma(N)} \gamma\left( N, \frac{N T_{AF}^{*}}{\sigma^{2}} \right) - Q_{N} \left( \sqrt{\frac{2 N |\alpha_{AF}|^{2}}{\sigma^{2}}}, \sqrt{\frac{2 N T_{AF}^{*}}{\sigma^{2}}} \right) \right], \label{BER_AFa} \\
p_{b}^{opt,AFb} &= \frac{1}{2} \left[ \frac{1}{\Gamma(N)} \gamma\left( N, \frac{N T_{AF}^{*}}{\sigma^{2}} \right) + Q_{N} \left( \sqrt{\frac{2 N |\alpha_{AF}|^{2}}{\sigma^{2}}}, \sqrt{\frac{2 N T_{AF}^{*}}{\sigma^{2}}} \right) \right]. \label{BER_AFb}
\end{align}
\end{subequations}

The exact BER expressions under the Gaussian-approximated and simple thresholds can be readily obtained by replacing the exact thresholds above with the solutions of (\ref{thm2}) and (\ref{simpleThreshold}) in the BER of the respective link in the above sets of equations. 

Given the presence of the Marcum-Q and gamma functions, it is unlikely that closed-form expressions for the outage probability exist. However, numerical results can be readily obtained using standard mathematical packages.

\subsection{Source Power Optimization under the DF Scheme}

Due to the passive nature of the backscatter relay communication, the performance of the DF system is likely to be limited by the relay-to-destination link. Therefore, a key question to be answered to ensure the optimal operation of the DF scheme is how much power the source should allocate to the first timeslot compared to the second timeslot. The question becomes particularly relevant when the source is subject to a power budget constraint.

We denote the total power budget of the source by $P_{S} \triangleq P_{S,1} + P_{S,2}$. The power allocation problem can be written as
\begin{equation}
\begin{aligned}
& \min_{P_{S,1}} & & p_{b}^{DF} \\
& \text{s.t.} & & P_{S,1} + P_{S,2} = P_{S}.	\label{P}
\end{aligned}
\end{equation}
Note that due to the incomplete gamma function and Marcum Q-function in (\ref{BER_DF1a})-(\ref{BER_DF2b}) and the threshold equations in (\ref{thm1b})-(\ref{thm1c}), there is no closed-form solution to (\ref{P}). However, by setting $P_{S,2} = P_{S} - P_{S,1}$, the optimal allocation can be determined by taking the first derivative of $P_{b}^{DF}$ with respect to $P_{S,1}$ and evaluating the roots. 

\begin{proposition}
An approximation of the optimal solution to (\ref{P}) is the solution to
\begin{align}
&\frac{N}{2} \left[  \frac{|\alpha_{1}|^{2}}{\sigma_{1}^{2}} Q_{N+1}^{-}\left( \frac{\sqrt{2N |\alpha_{1}^{'}|^{2} (P_{S} - P_{S,1})}}{\sigma_{1}^{2}}, \sqrt{\frac{2 N T_{D}^{*}}{\sigma_{1}^{2}}} \right) \right. \nonumber \\
	& \quad \left. - \frac{|\alpha_{0}|^{2}}{\sigma_{0}^{2}} Q_{N+1}^{-}\left( \frac{\sqrt{2N |\alpha_{0}^{'}|^{2} (P_{S} - P_{S,1})}}{\sigma_{0}^{2}}, \sqrt{\frac{2 N T_{D}^{*}}{\sigma_{0}^{2}}} \right) \right. \nonumber \\
	& \quad \left. - \frac{1}{\sigma^{2}} Q_{N+1}^{-}\left( \sqrt{\frac{2 N P_{S,1} |h_{SR}|^{2}}{\sigma^{2}}}, \sqrt{\frac{2 N T_{R}^{*}}{\sigma^{2}}} \right) \right] = 0, \label{P_sol}
\end{align}
where the product term in (\ref{twoLinkBER}) is dropped, and have defined $Q_{M}^{-}(a, b) \triangleq Q_{M}(a, b) - Q_{M-1}(a, b)$, and $\alpha_{i}^{'} = \eta h_{SR} h_{RD} B[n]$ for $i \in \{0, 1\}$.
\end{proposition}

If there exist multiple roots to (\ref{P_sol}), then the optimal power allocation can be determined by substituting each root into the BER expression in (\ref{twoLinkBER}) to determine the root resulting in the lowest BER. It should be noted that the optimal power allocation changes with $N$, which determines the effective data rate. In the case where a large portion of the power budget is allocated to the source-to-relay link, the relay-to-destination link will perform poorly, which affects the end-to-end performance. This is due to the performance of backscatter devices being highly dependent on the incident signal power. At the other extreme, when a large portion of power is assigned to the relay-to-destination link, that link will perform well at the expense of the source-to-relay link. As  a result, end-to-end BER performance is still poor. Intuitively, there exists an optimal power allocation, arising from the case where the BER of the two individual links are roughly equal. This allocation occurs at a point between the two mentioned extremes.


\section{Numerical Results}

In this section, we numerically evaluate the performance of the proposed DF and AF backscatter relaying schemes. The set of system parameters used to obtain the numerical results are provided in Table~I. Based on works such as \cite{Amb16, yang2018modulation}, we consider a BER of $10^{-2}$ to be acceptable performance.

\begin{table}
\caption{List of System Parameters.}
\label{parameters}
\setlength{\tabcolsep}{3pt}
\centering
\begin{tabular}{|c|c|}
\hline
\textbf{Parameter} & \textbf{Value} \\
\hline\hline
Carrier frequency, $f_{c}$ & $915$ MHz \\
\hline
Source-to-relay distance, $d_{SR}$ & $15$ m \\
\hline
Relay-to-destination distance, $d_{RD}$ & $15$ m \\
\hline
Path loss exponent, $\gamma$ & $2.5$ \\
\hline
Rician $K$-factor & $4$ \\
\hline
Transmit antenna gain, $G_{t}$ & $6$ dB \\
\hline
Relay antenna gain, $G_{r}$ & $1.5$ dB \\
\hline
Noise powers, $\{P_{w,R}, P_{w,D}\}$ & $\{-110, -110\}$ dBm \\
\hline
Antenna structural mode at relay, $A$ & $0.6047 + j0.5042$ \cite{bistatic} \\
\hline
Reflection coefficients at relay, $\{\Gamma_{0}, \Gamma_{1}\}$ & $\{\approx A, -|A| / A\}$ \\
\hline
Backscatter switching loss coefficient, $\eta$ & $-1.1$ dB \cite{passive} \\
\hline
Samples per source symbol, $N$ & $25$ \\
\hline
\end{tabular}
\end{table}

First, we present a case study based on the system parameters in Table~I to demonstrate the blind spot scenario. Suppose a sensor network is installed on the external perimeter of a building with the purpose of performing structural monitoring, in addition to monitoring environmental variables. The direct link through the building, comprised of concrete supports, experiences severe attenuation of around $35$ dB for each obstacle of $0.3$ m thickness at $1$ GHz \cite{nist}, resulting in overall attenuation of around $140$ dB when three such obstacles are considered, in addition to the reference path loss of $32$ dB. The relay link, however, experiences $64$ dB reference path loss due to the two links, in addition to $59$ dB of combined path loss over both links. In this case, the relay link is around $1.5$ orders of magnitude stronger than the direct link. Hence, the contribution of the direct link can be considered as negligible.

\begin{figure}
	\centering
	\subfigure[DF scheme with optimized power allocation]
	{
		\centering
		\includegraphics[width=3.5in]{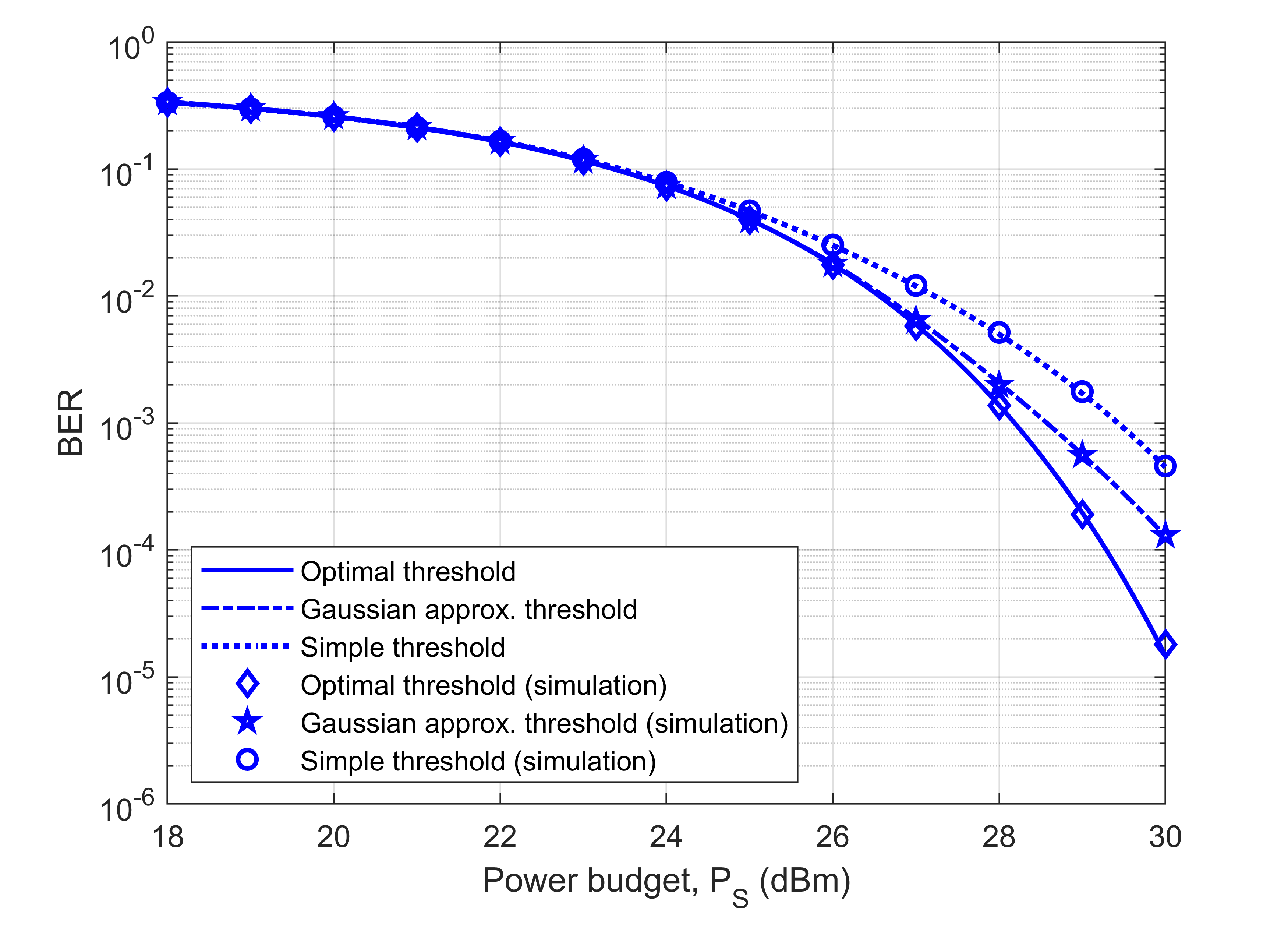}
	}
	\vfill
	\subfigure[AF scheme]
	{
		\centering
		\includegraphics[width=3.5in]{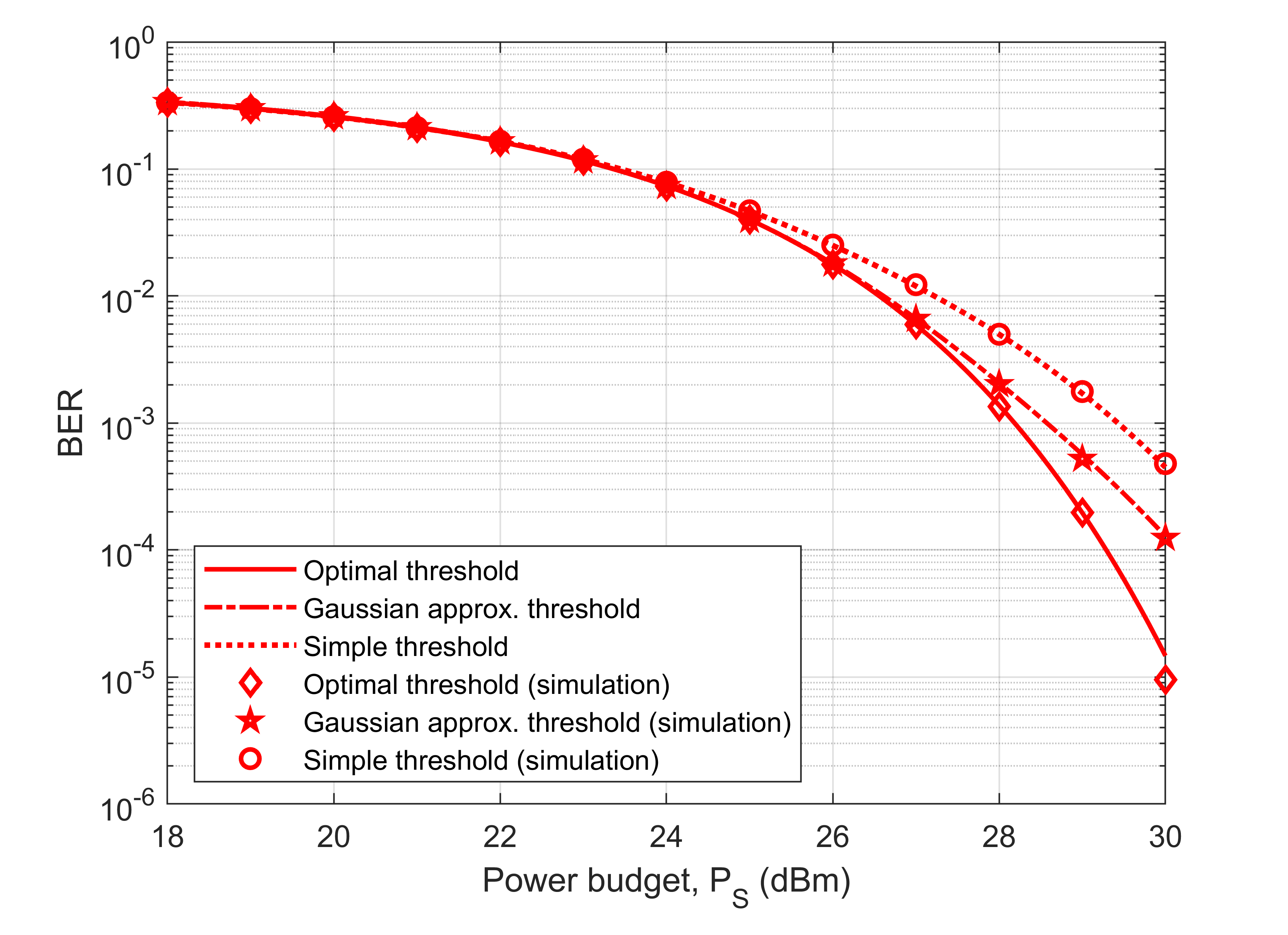}
	}
	\label{fig:2}
	\caption{BER at destination vs. source power budget.}
\end{figure}

\subsection{Performance under Optimal, Gaussian-Approximated and Simple Thresholds}

To provide a performance comparison of the three thresholds, we consider the BER over one channel coherence period with unit channel gains. Fig. 2 shows the BER of both DF and AF schemes, with the interference power at backscatter relay and destination taking on values of $-70$ and $-85$ dBm, respectively. For the DF scheme, the power allocation problem in (\ref{P}) is solved for each power budget value. The resulting BER is obtained by substituting the optimal allocation into the threshold equations in (\ref{thm1a})-(\ref{simpleThreshold}). The thresholds are then substituted into the BER equations in (\ref{BER_DF1a})-(\ref{BER_AFb}). The analytical performance of the optimal, Gaussian-approximated and simple thresholds is compared with simulation results, which are obtained by averaging over $2000$ iterations for each power budget value, where $1000$ source symbols are transmitted per iteration. We note that for both DF and AF schemes, the simulated BER results match exactly with analytical results obtained from the threshold expressions in Section IV and BER expressions in Section V. Hence, we will only present analytical results hereafter.

The performance under the Gaussian-approximated threshold closely resembles that of the optimal threshold for BER up to $10^{-3}$, and begins to exhibit some minor performance degradation thereafter. In the high power budget regime, i.e. when the power budget is around $30$ dBm, using the CLT to approximate the test statistic distribution is no longer suitable and results in considerably suboptimal performance. However, the computation of the optimal threshold is highly complex. As a result, it is reasonable to suggest that the Gaussian threshold performs satisfactorily for moderate BER requirements, given device-level complexity considerations. Unless otherwise noted, in the following subsections we present results using the Gaussian-approximated threshold only.

Compared to the Gaussian-approximated threshold, the simple threshold begins to exhibit worse performance at BER values below $10^{-1}$, with the performance gap increasing to around $1$ dB in the high power budget regime. Again, this highlights the trade-off between performance and complexity.

\begin{figure}
	\centering
	\subfigure[Varying $P_{I,R}$ with $P_{I,D} = -90$ dBm]
	{
		\centering
		\includegraphics[width=3.5in]{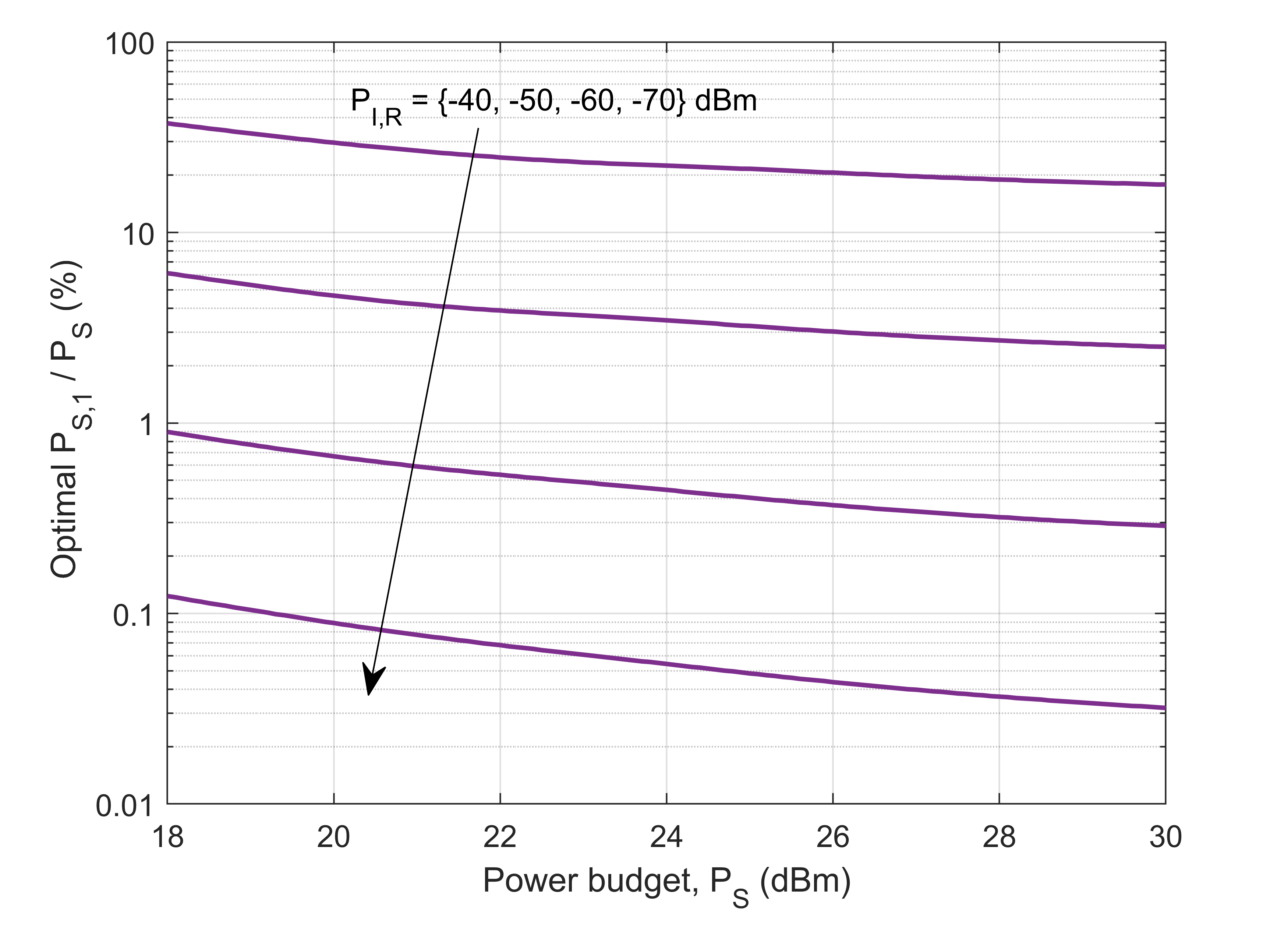}
	} \vfill
	\subfigure[Varying $P_{I,D}$ with $P_{I,R} = -70$ dBm]
	{
		\centering
		\includegraphics[width=3.5in]{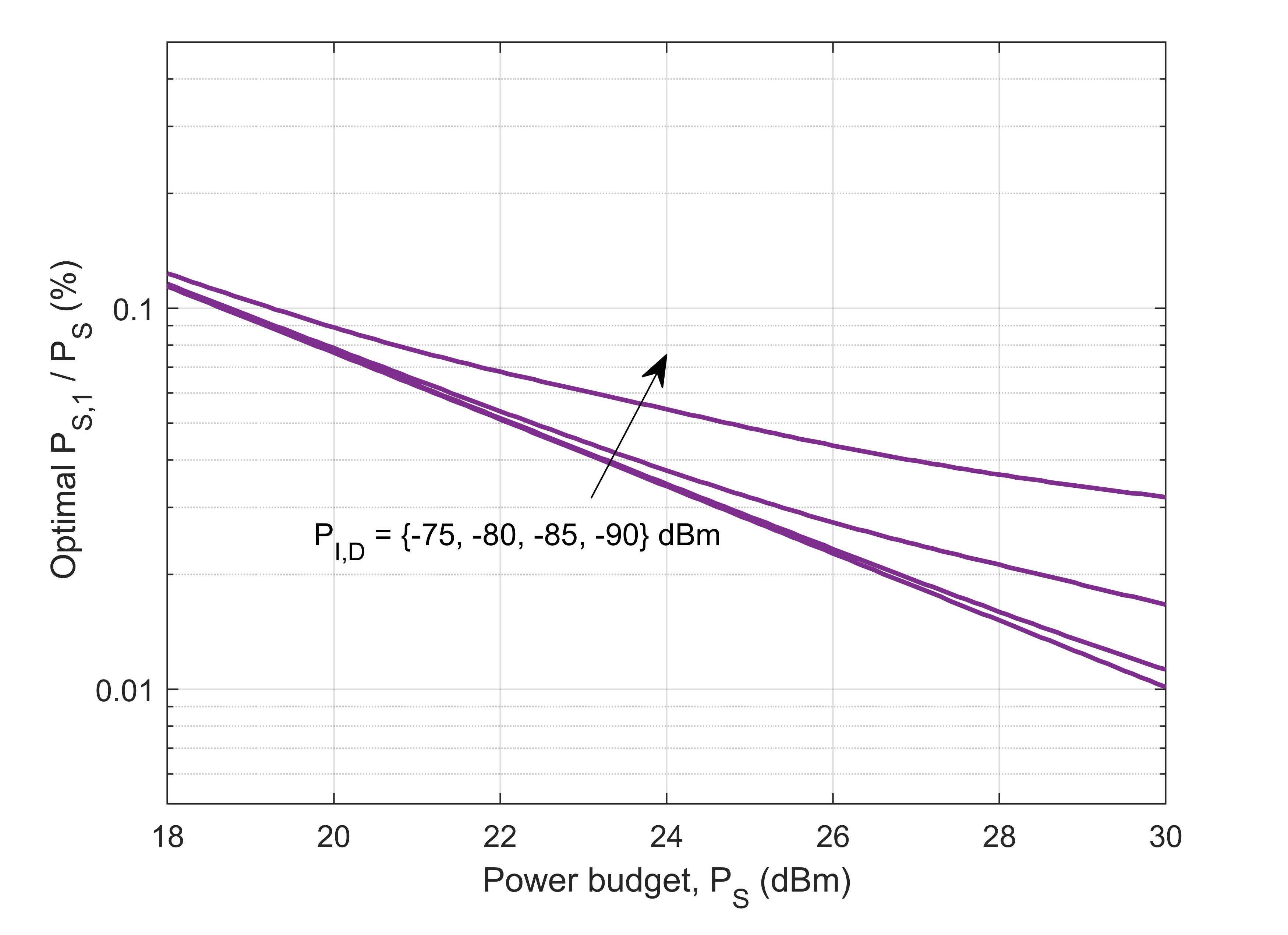}
	}
	\label{fig:3}
\caption{Optimal power allocation for the first timeslot as a percentage of total power budget.}
\end{figure}

\begin{figure}
	\centering
	\includegraphics[width=3.5in]{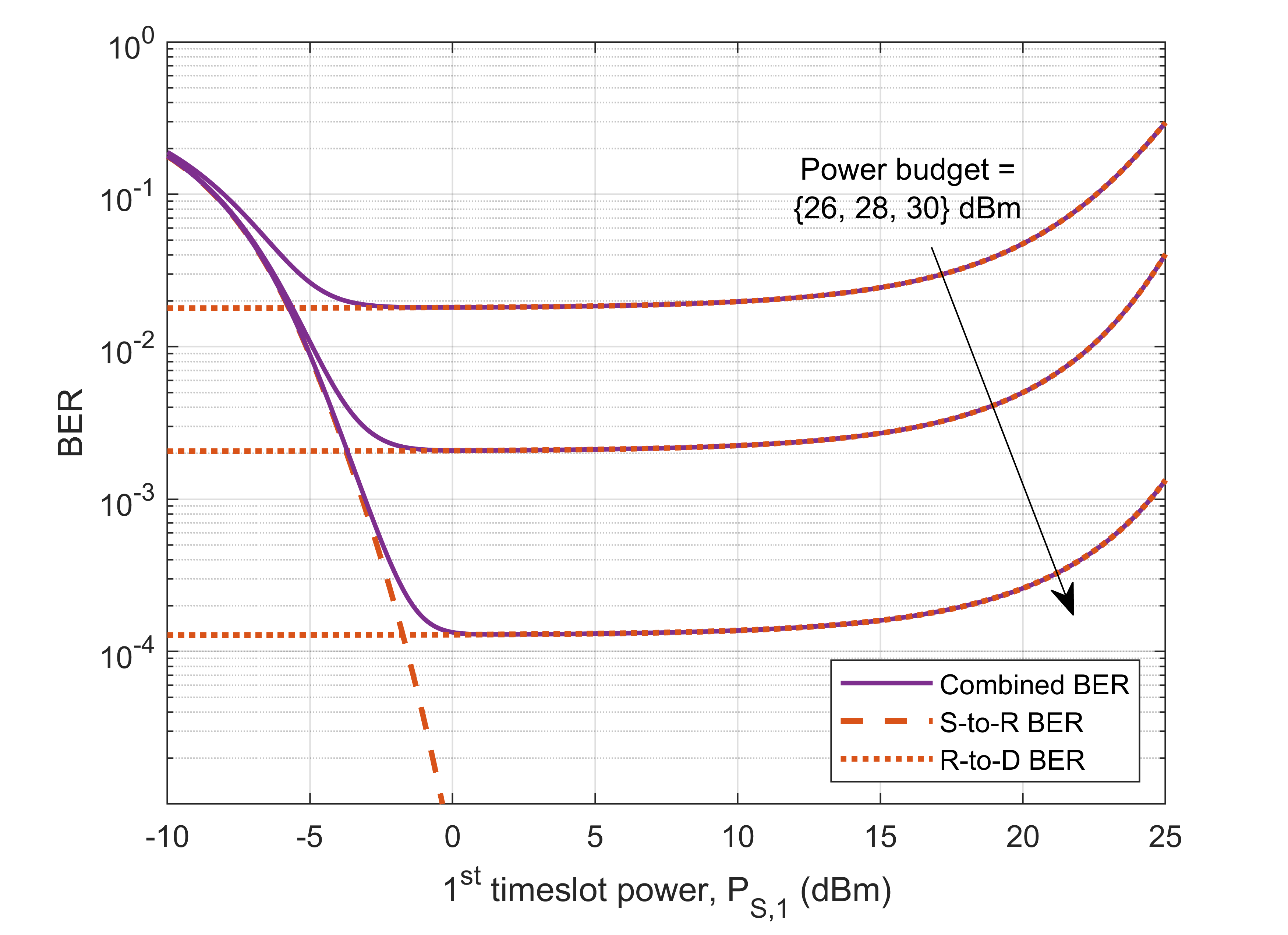}
	\caption{Combined BER for end-to-end transmission under the DF scheme, in addition to BERs for the source-to-relay and relay-to-destination links vs. power allocation to the first timeslot.}
	\label{fig:4}
\end{figure}

\subsection{Optimal Power Allocation under the DF Scheme}

Fig. 3 shows the optimal percentage of power allocated to the first timeslot under the DF scheme, with the interference power at the relay ($P_{I,R}$) and the destination ($P_{I,D}$) held constant at the baseline values of $-70$ and $-90$ dBm, respectively. It is evident that the optimal percentage is far below $50\%$ for the range of $P_{I,R}$ values, meaning the majority of power is assigned to the second timeslot. This is because the backscattered information signal received at the destination is typically much weaker than the received interference signal. To boost the backscattered signal strength, the source needs to transmit a high-power CW signal. This implies that the relay-to-destination link is the limiting link under most circumstances, and hence requires larger power allocation. This observation is consistent with the trend that the proportion of power allocated to the first timeslot reduces as the source power budget increases, regardless of interference levels.

Fig. 4 plots the BER under the DF scheme for a range of power budgets, in addition to the individual contributions of the source-to-relay and relay-to-destination link BER. For illustration, we set the interference powers to $P_{I,R} = -60$ dBm and $P_{I,D} = -85$ dBm and assume unit channel gain. One can observe that the optimal value of $P_{S,1}$ occurs around the point where $p_{b}^{(1)} = p_{b}^{(2)}$, confirming intuition in Section V. Past this point, the combined BER is largely dependent on $p_{b}^{(2)}$. For small power budgets, even if the majority of the power is allocated to the second timeslot, the overall performance is ultimately limited by the BER of the source-to-relay link.

\subsection{Comparison Between DF and AF Schemes}

We then compare the performance of the DF and AF schemes in terms of their long-term outage probability, under scenarios where the interference power at either the relay or destination is varied. We also consider different values of the number of samples per symbol and the relay aperture size, which are fixed once chosen; in addition to a case of imperfect impedance matching at the relay. For the remaining results, the BER threshold for outage is $10^{-2}$, and the outage probability is obtained by averaging over $5000$ channel coherence periods.

\begin{figure}
	\centering
	\subfigure[DF scheme]
	{
		\centering
		\includegraphics[width=3.5in]{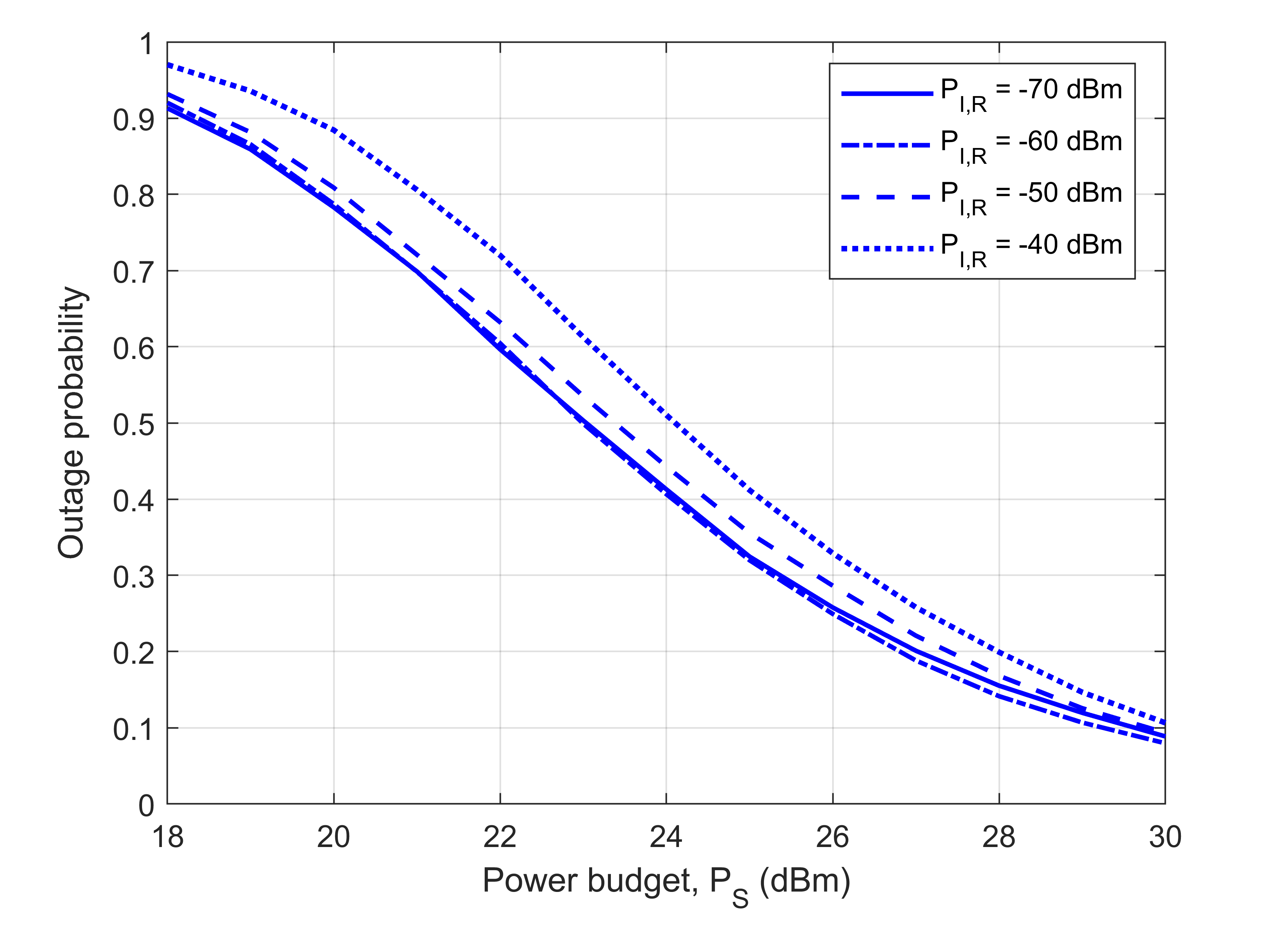}
	} \vfill
	\subfigure[AF scheme]
	{
		\centering
		\includegraphics[width=3.5in]{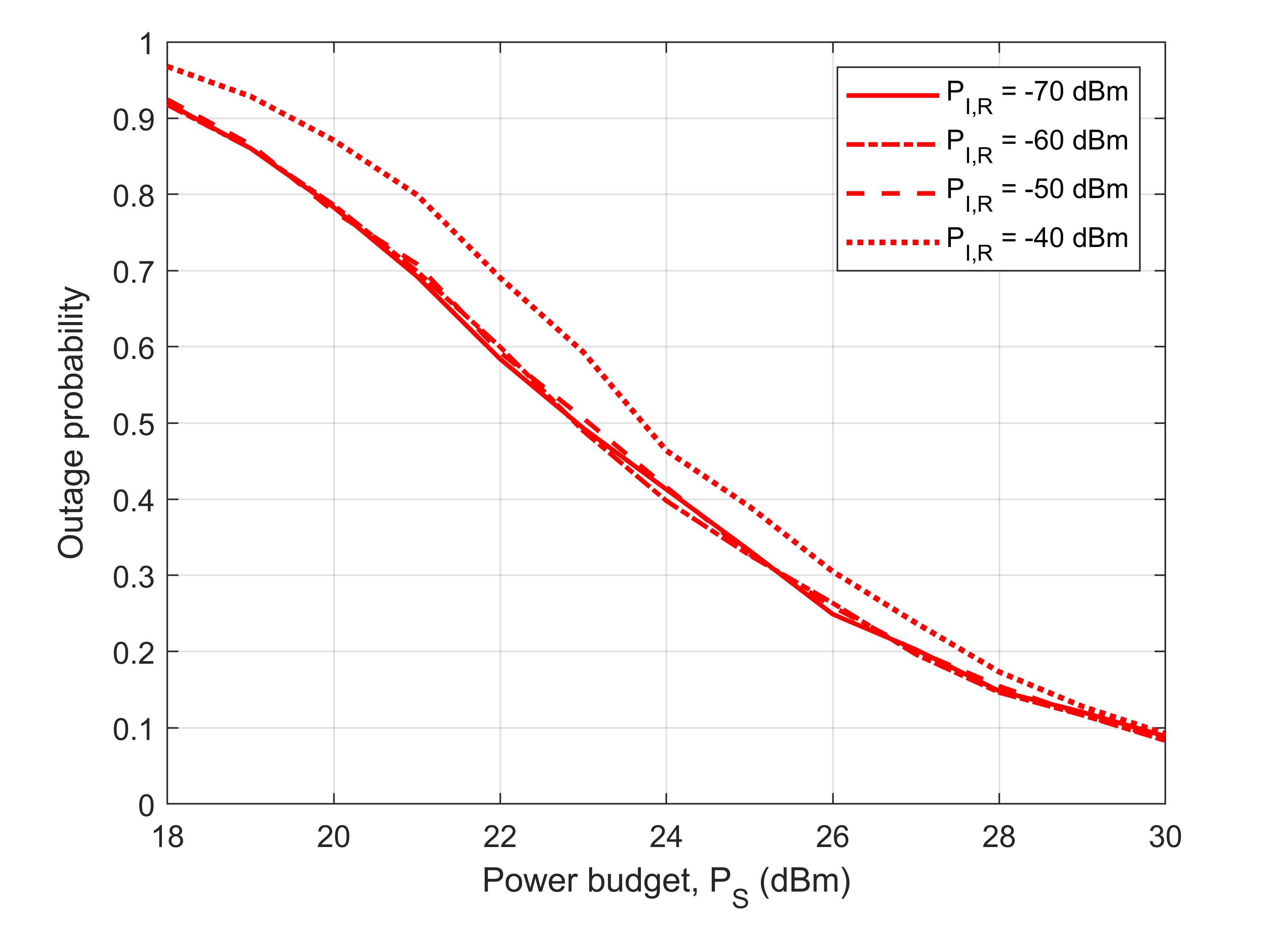}
	}
	\label{fig:5}
	\caption{Comparison between the DF and AF schemes under various interference powers at the relay, with $P_{I,D} = -90$ dBm.}
\end{figure}

Fig. 5 compares the outage probability of the DF and AF schemes where the interference power at the relay is varied, holding the interference power at the destination constant. For each interference level, both the DF and AF schemes perform similarly, indicating very little performance loss for the DF scheme when the reflection coefficients are carefully selected. Notably, however, even as the relay experiences stronger interference, the outage probabilities under both schemes only show minor deterioration. This highlights the robustness of the power allocation for the DF scheme, which is able to tolerate a high level of interference by only allocating a small power budget to the source-to-relay link before significantly increased outage behavior is observed.

\begin{figure}
	\centering
	\subfigure[DF scheme]
	{
		\centering
		\includegraphics[width=3.5in]{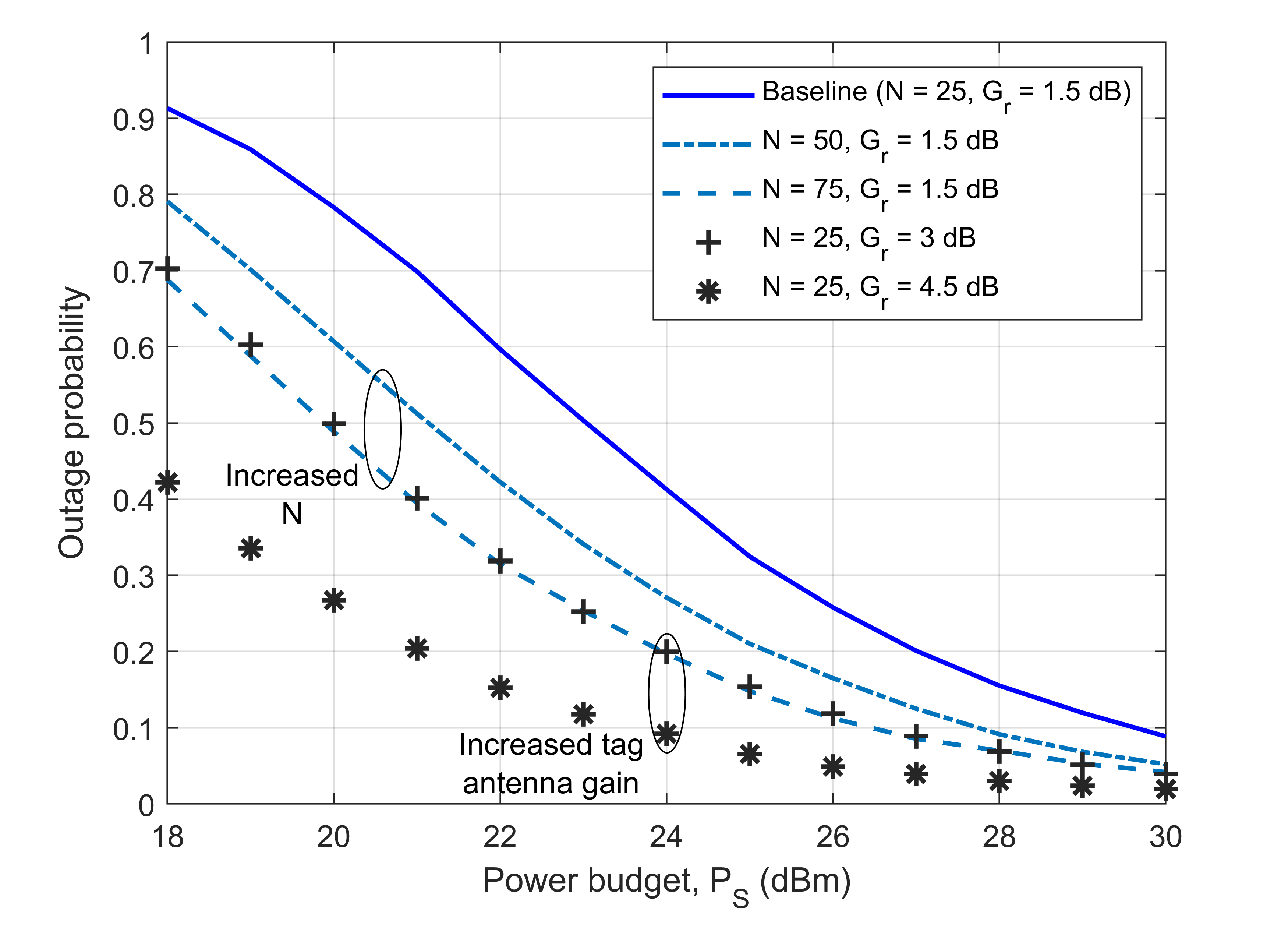}
	} \vfill
	\subfigure[AF scheme]
	{
		\centering
		\includegraphics[width=3.5in]{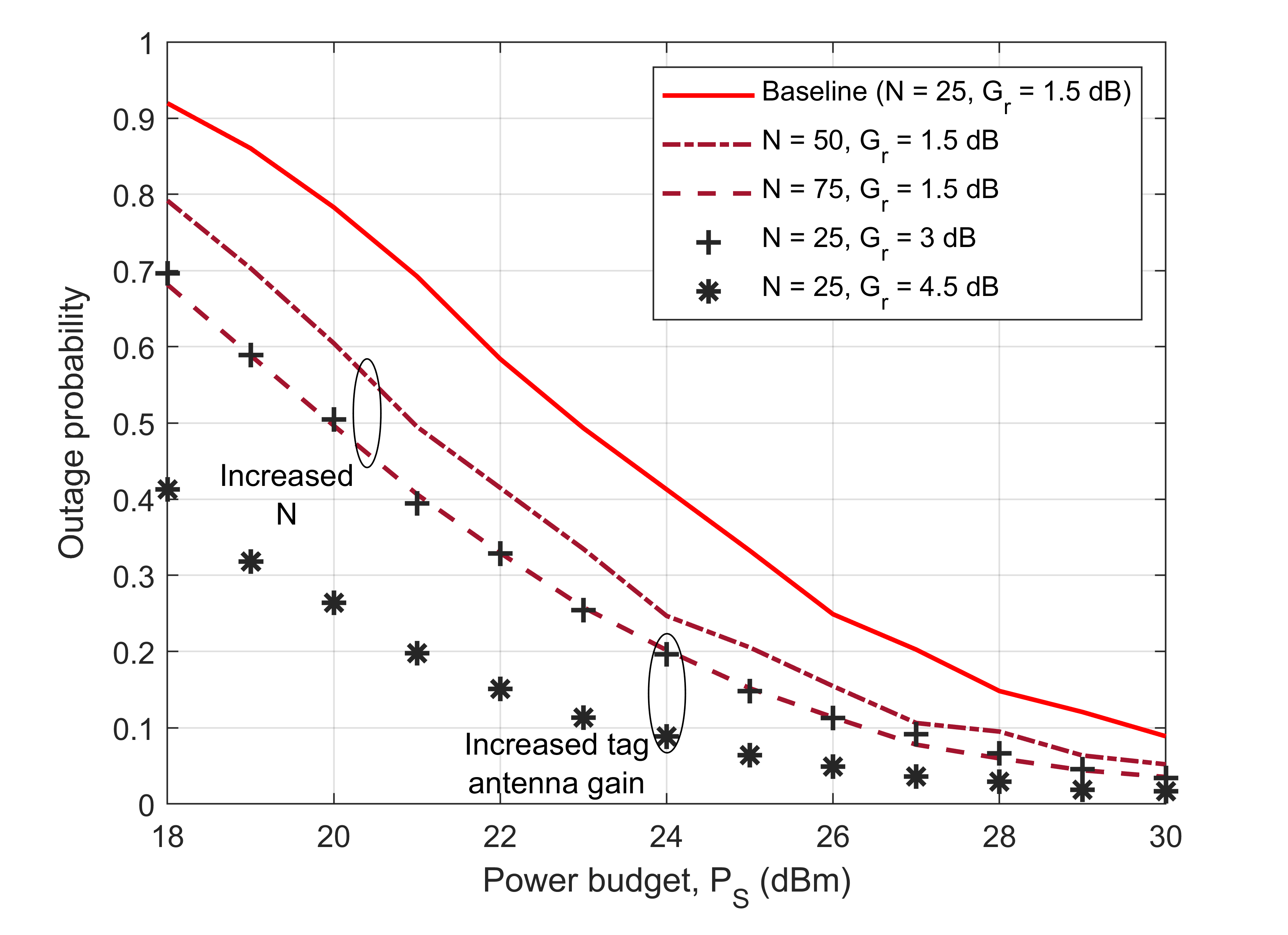}
	}
	\label{fig:6}
	\caption{Comparison between the DF and AF schemes under various antenna gains and number of samples per symbol, with the baseline interference powers being $P_{I,R} = -70$ dBm and $P_{I,D} = -90$ dBm.}
\end{figure}

Fig. 6 highlights the effects of increasing the number of samples per symbol $N$ and the antenna aperture area at the relay. Note that the changing value of $N$ here is solely to highlight the trade-offs between the data rate and the reliability of the system, with $N$ being a design parameter that is fixed for all transmissions once chosen. Specifically, an increase in $N$ can be interpreted as a reduction in data rate (keeping the sampling rate the same), in order to gain reliability. It is observed that a two-fold ($3$ dB) increase from $N = 25$ to $N = 50$ results in a $2$ dB improvement in the outage probability, and a three-fold ($4.77$ dB) increase to $N = 75$ leads to a $3$ dB improvement. However, linear scaling behavior is observed when the size of the relay aperture is increased, by a factor of $2$ ($3$ dB) and $4$ ($6$ dB). Note that doubling the effective antenna aperture is equivalent to doubling the antenna gain. Therefore, increasing the size of the antenna is more effective than increasing $N$, as evidenced by the $3$ dB vs. $1.8$ dB improvement in the outage probability when doubling the aperture size compared to doubling the symbol time.

\begin{figure}
	\centering
	\subfigure[DF scheme]
	{
		\centering
		\includegraphics[width=3.5in]{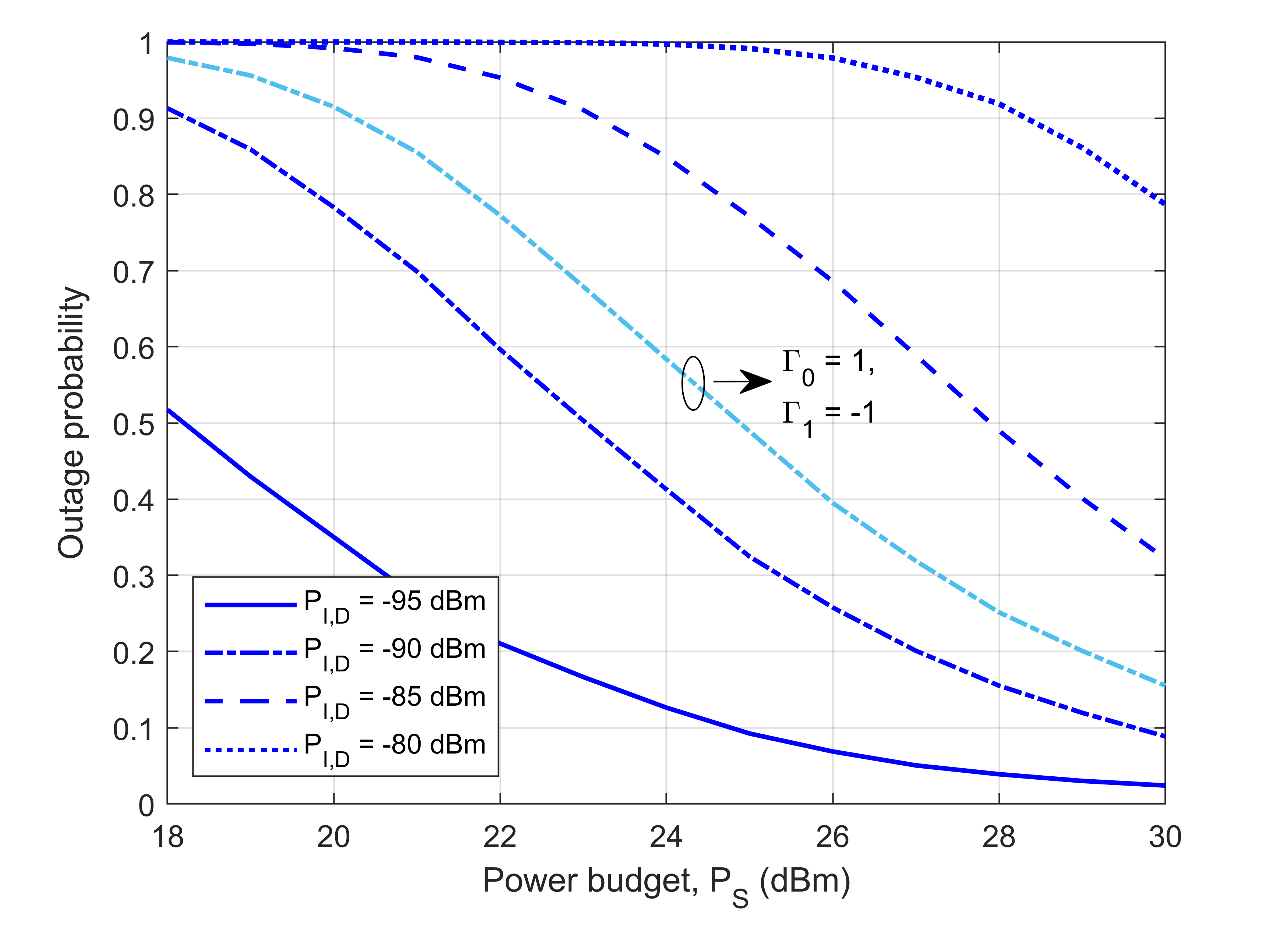}
	} \vfill
	\subfigure[AF scheme]
	{
		\centering
		\includegraphics[width=3.5in]{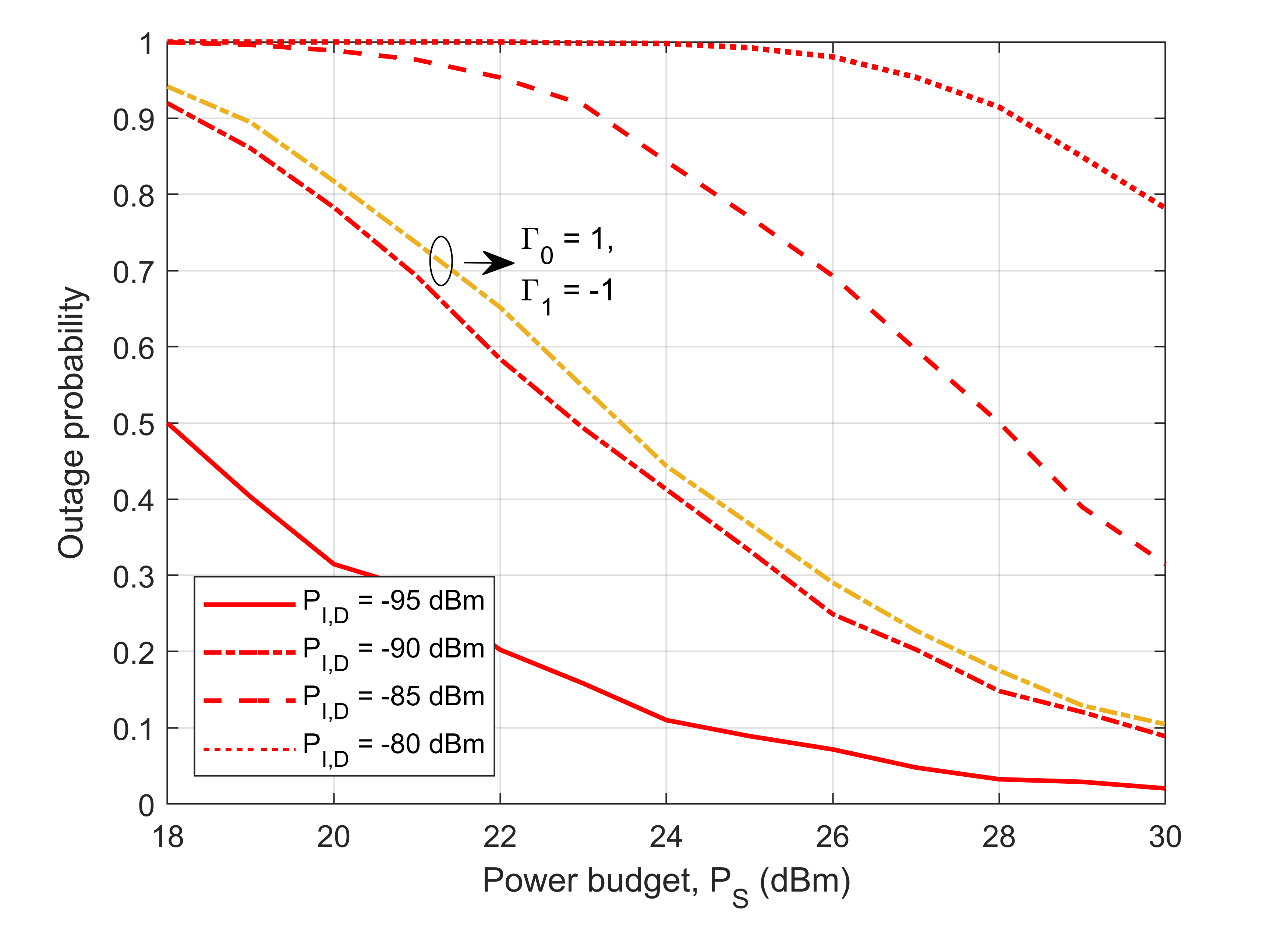}
	}
	\label{fig:7}
	\caption{Comparison between the DF and AF schemes under various interference powers at the destination, with $P_{I,R} = -70$ dBm, in addition to the case where the reflection coefficients at the relay are modified.}
\end{figure}

In Fig. 7, when the interference experienced by the relay is held constant, the outage probability performance only becomes acceptable when $P_{I,D}$ is suppressed to be close to $-85$ dBm or less. This reaffirms the fact that interference at the destination is the main determinant of detection performance. These properties are unique to the backscatter relay system and could give rise to applications where interference mitigation is of importance, or where interference could be judiciously used to enhance performance.

We also illustrate the impact of the choice of reflection coefficient at the backscatter relay in Fig. 7. Note that the values of $\Gamma_{0}$ and $\Gamma_{1}$, given in Table~I and used in all previous results, represent perfect OOK modulation, i.e. the antenna and load impedances at the relay are perfectly matched. In practice, perfect impedance matching may not be possible. The antenna structural mode may vary between backscatter devices due to small variations in the antenna material; or the impedances may not be optimally designed for specific applications (e.g. semi-passive transmissions). Here, we consider the latter scenario with different values for $\Gamma_{0}$ and $\Gamma_{1}$. 

For this study, the reflection coefficients used at the relay are $\Gamma_{0} = 1$ and $\Gamma_{1} = -1$, which are the values used in \cite{bistatic}, under the interference powers of $P_{I,R} = -70$ dBm and $P_{I,D} = -90$ dBm. It is observed that the AF scheme outperforms the DF scheme by a considerable margin. This is due to the fact that under the AF scheme, the relay uses the impedance that results in higher average signal power received at the destination, regardless of the bit transmitted by the source; whereas with DF, both relay impedances are used, leading to a lower signal power. Here, AF outperforms DF over the range of power budgets, and the advantage of AF upon DF is around $1.5$ dB when $P_{I,D}$ becomes small. This suggests that deviating from perfect OOK modulation by using other reflection coefficients drastically degrades the performance of the DF scheme, while only resulting in moderate degradation for the AF scheme. As a result, AF outperforms DF in relative terms. Nonetheless, comparing with the results from perfect impedance matching in the same figure, it can be seen that a simple choice of reflection coefficients results in performance degradation for both schemes.

\subsection{Summary of Results}
Our main findings on the backscatter relay system are as follows:

\begin{itemize}
\item For the DF scheme, the optimal allocation of source power budget assigns a much larger proportion of power to the second timeslot to support the relay-to-destination link. Improper choices of power allocation can result in significant performance degradation.
\item For the DF scheme, interference received by the backscatter relay has minor impact on the end-to-end outage probability, as long as the source power budget is sufficient to maintain the BER performance of both source-to-relay and relay-to-destination links.
\item Due to its increased complexity, the performance of the DF scheme is inferior to the AF scheme when the modulation deviates from perfect OOK. However, it is always possible to improve the performance of the DF scheme by extending the symbol period in terms of signal samples, or to incorporate error correction codes.
\end{itemize}


\section{Conclusion and Future Work}
In this paper, the performance of the DF and AF backscatter relaying schemes in the presence of ambient interference was examined. The exact and approximate distributions of the average received signal power at both relay and destination were derived, which enabled us to formulate the corresponding test statistics required to obtain the detection thresholds for energy-based detection. The optimal power allocation at the source under the DF scheme was also studied. Several notable results were summarized in Section VI-D.

This paper provides a fundamental analysis for the relay use case of backscatter devices; as such, a number of scenarios remain to be studied. The assumption used in this work that the circuit energy consumption at the relay is negligible can be replaced with a realistic circuit constraint. The effect of error correction coding on BER performance can be examined, accounting for energy consumption. Finally, the multi-hop backscatter relaying with mode selection between the DF and AF schemes can be considered.


\appendices
\section{Proof of Proposition 1}
We expand the expression for the test statistic under the DF scheme at the relay into its real and imaginary components, and characterize its exact distribution. For $\psi^{DF,R}_{0}$, we have
\begin{equation}
\psi^{DF,R}_{0} = \frac{1}{N} \sum_{n=0}^{N-1} \left( \sqrt{P_{I,R}} z_{r}[n] + w_{r}[n] \right)^{2} + \frac{1}{N} \sum_{n=0}^{N-1} \left( \sqrt{P_{I,R}} z_{i}[n] + w_{i}[n] \right)^{2}, \label{appA_2}
\end{equation}
where $z_{r}[n], z_{i}[n]$ and $w_{r}[n], w_{i}[n]$ denote the real and imaginary components of $z_{R,1}[n]$ and $w_{R}[n]$, respectively, here and elsewhere in the appendices. Note that (\ref{appA_2}) represents the sum of $2N$ squared zero-mean Gaussian random variables, each with variance $\varsigma^{2} = \frac{1}{2} \left(P_{I,R} + P_{w,R}\right)$. Denoting each Gaussian random variable by $X \sim \mathcal{N}(0, \varsigma^{2})$, it follows that $\frac{X}{\varsigma} \sim \mathcal{N}(0, 1)$. Squaring both sides gives $\frac{X^{2}}{\varsigma^{2}} \sim \chi^{2}(1) = \Gamma(\frac{1}{2}, 2)$, that is, a gamma distribution. By the scaling properties of gamma random variables, we obtain $X \sim \Gamma(\frac{1}{2}, 2\varsigma^{2})$ after rearranging. Summing $2N$ i.i.d. gamma random variables and multiplying by the factor $\frac{1}{N}$ gives $\psi^{DF,R}_{0} \sim \Gamma\left(N, \frac{2 \varsigma^{2}}{N}\right)$, which is equivalent to the representation in (\ref{prop1a}).

The derivation is similar for $\psi^{DF,R}_{1}$, and is based on the proof in \cite{exactBER}. We let $\Psi_{1} = \frac{2 N}{\varsigma^{2}} \psi_{1}$, which accounts for the variances of the Gaussian random variables in the following equation:
\begin{equation}
\Psi_{1} = \frac{2}{\varsigma^{2}} \sum_{n=0}^{N-1} \left( \sqrt{P_{S,1}} h_{r} + \sqrt{P_{I,R}} z_{r}[n] + w_{r}[n] \right)^{2} + \frac{2}{\varsigma^{2}} \sum_{n=0}^{N-1} \left( \sqrt{P_{S,1}} h_{i} + \sqrt{P_{I,R}} z_{i}[n] + w_{i}[n] \right)^{2},	\label{appA_4}
\end{equation}
where $h_{r}$ and $h_{i}$ are the real and imaginary components of the channel coefficient $h_{SR}$. Again, there are $2N$ squares of i.i.d. Gaussian random variables each with nonzero mean and variance $\varsigma^{2}$ in (\ref{appA_4}). When summed, the result is a NC-$\chi^{2}$ random variable with $2N$ degrees of freedom. Note that the same $\varsigma$ is used in both (\ref{appA_2}) and (\ref{appA_4}), as $P_{S,1}$, $h_{r}$ and $h_{i}$ are all constants within each channel coherence period. The noncentrality parameter $\lambda$ is dependent on the means of the Gaussian random variables in (\ref{appA_4}):
\begin{equation}
\lambda = \frac{2}{\varsigma^{2}} \sum_{n=0}^{N-1} \left( \sqrt{P_{S,1}} h_{SR} \right)^{2} = \frac{2 N P_{S,1} |h_{SR}|^{2}}{\varsigma^{2}}.	\label{appA_5}
\end{equation}
Note that $\psi^{DF,R}_{1}$ is obtained by scaling $\Psi_{1}$ by $\frac{\sigma^{2}}{2 N}$; however, this does not affect the noncentrality parameter in (\ref{appA_5}). Given that there are $2N$ degrees of freedom for the NC-$\chi^{2}$ random variable, we obtain the representation in (\ref{prop1b}).


\section{Proof of Theorem 1}
We present the derivation for (\ref{thm1a})-(\ref{thm1b}). Note that the modified Bessel function of the first kind $I_{\nu}(\cdot)$ can be written in the following integral form for integer values of $\nu$:
\begin{equation}
I_{\nu}(x) = \frac{1}{\pi} \int_{0}^{\pi} \exp\left( x \cos(\theta) \right) \cos(v \theta)\  \textrm{d}\theta.	\label{appB_1}
\end{equation}
Therefore, for the DF scheme at the relay, equating the pdfs for the two possible symbols gives the following:
\begin{multline}
\frac{1}{\Gamma(n) \left( \frac{\sigma_{DF}^{2}}{N} \right)^{N}} x^{N-1} \exp \left( -\frac{N x}{\sigma_{DF}^{2}} \right) = \frac{N}{\pi \sigma_{DF}^{2}} \exp \left( -\frac{N}{\sigma_{DF}^{2}} \left( x + P_{S,1} |h_{SR}|^{2} \right) \right) \left( \frac{\sigma_{DF}^{2} x}{P_{S,1} |h_{SR}|^{2}} \right)^{\frac{N-1}{2}} \\ \times \int_{0}^{\pi} \exp \left( \frac{2N}{\sigma_{DF}^{2}} \sqrt{P_{S,1} |h_{SR}|^{2} x} \cos(\theta) \right) \cos(N-1)\theta \ \textrm{d}\theta.	\label{appB_2}
\end{multline}
Rearranging and simplifying (\ref{appB_2}) results in the expression in (\ref{thm1a}).
For the DF scheme at the destination, the two pdfs given in (\ref{prop3}) are set equal to each other (we denote the integrals in $f_{\psi_{0}}(x)$ and $f_{\psi_{1}}(x)$ as $I_{0}$ and $I_{1}$, respectively):
\begin{multline}
\frac{N}{\sigma_{0}^{2}} \exp \left( -\frac{N}{\sigma_{0}^{2}} \left( x + |\alpha_{DF,0}|^{2} \right) \right) \left( \frac{\sigma_{0}^{2} x}{|\alpha_{DF,0}|^{2}} \right)^{\frac{N-1}{2}} I_{0} \\
= \frac{N}{\sigma_{1}^{2}} \exp \left( -\frac{N}{\sigma_{1}^{2}} \left( x + |\alpha_{DF,1}|^{2} \right) \right) \left( \frac{\sigma_{1}^{2} x}{|\alpha_{DF,1}|^{2}} \right)^{\frac{N-1}{2}} I_{1}.	\label{appB_3}
\end{multline}
When simplified, (\ref{appB_3}) can be written as
\begin{equation}
\frac{\sigma_{1}^{2}}{\sigma_{0}^{2}} \exp \left( \left( \frac{N}{\sigma_{1}^{2}} - \frac{N}{\sigma_{0}^{2}} \right) x + \left( \frac{N |\alpha_{DF,1}|^{2}}{\sigma_{1}^{2}} - \frac{N |\alpha_{DF,0}|^{2}}{\sigma_{0}^{2}} \right) \right) \left( \frac{|\alpha_{DF,1}|^{2}}{|\alpha_{DF,0}|^{2}} \right)^{\frac{N-1}{2}} I_{0} = I_{1}.	\label{appB_4}
\end{equation}
Note that the only difference between $|\alpha_{DF,0}|^{2}$ and $|\alpha_{DF,1}|^{2}$ is the relay baseband signal, which are $B_{0}$ and $B_{1}$ for the two terms, respectively. From this, we arrive at the expression in (\ref{thm1b}). The result in (\ref{thm1c}) can be obtained analogously by following the steps taken to derive (\ref{thm1a}). Due to the lack of a closed form for integral in (\ref{appB_1}), there are no exact closed-form expressions for the thresholds in (\ref{thm1a})-(\ref{thm1c}). However, the thresholds can be numerically computed via standard mathematical packages.


\section{Proof of Theorem 2}
The proof is non-trivial, as we need to account for all values of $\mu_{0}$, $\mu_{1}$ and $\hat{\sigma}_{0}$, $\hat{\sigma}_{1}$ when the exact maximum-likelihood (ML) detection boundary is unknown. When the means and variances of the test statistics are available at a receiver (either relay or destination), the Gaussian-approximated pdf of any pair of $\psi_{0}$ and $\psi_{1}$ from Section IV-A to IV-C are given by
\begin{subequations}
\begin{align}
f_{\psi_{0}}(x) &= \frac{1}{\sqrt{2\pi \hat{\sigma}_{0}^{2}}} \exp\left(-\frac{(x-\mu_{0})^{2}}{2\hat{\sigma}_{0}^{2}}\right), \\
f_{\psi_{1}}(x) &= \frac{1}{\sqrt{2\pi \hat{\sigma}_{1}^{2}}} \exp\left(-\frac{(x-\mu_{1})^{2}}{2\hat{\sigma}_{1}^{2}}\right).	\label{appC_1}
\end{align}
\end{subequations}
The equation $f_{\psi_{0}}(x) = f_{\psi_{1}}(x)$ can be simplified to give
\begin{equation}
\left( \hat{\sigma}_{1}^{2} - \hat{\sigma}_{0}^{2} \right) x^2 + 2 \left( \hat{\sigma}_{0}^{2} \mu_{1} - \hat{\sigma}_{1}^{2} \mu_{0} \right) x + \left( \hat{\sigma}_{1}^{2} \mu_{0}^{2} - \hat{\sigma}_{0}^{2} \mu_{1}^{2} - 2 \hat{\sigma}_{0}^{2} \hat{\sigma}_{1}^{2} \ \textrm{ln}\left( \frac{\hat{\sigma}_{1}}{\hat{\sigma}_{0}} \right) \right) = 0.	\label{appC_2}
\end{equation}
When $|B_{1}| > |B_{0}|$, it follows that $\mu_{1} > \mu_{0}$ and $\hat{\sigma}_{1}^{2} > \hat{\sigma}_{0}^{2}$, and that there are exactly two solutions. Solving (\ref{appC_2}) gives the result in (\ref{thm2}). In the following, we present the conditions under which $T_{G}$ takes the solution with the positive sign (i.e. the positive solution). The conditions for the negative solution can be obtained analogously.

Examining the solution expressions in (\ref{thm2}), we see that the square root term always takes on a positive value: $\hat{\sigma}_{0}^{2} \hat{\sigma}_{1}^{2}$ is always positive; $\left(\mu_{0} - \mu_{1}\right)^{2}$ is always positive; $\hat{\sigma}_{0}^{2} - \hat{\sigma}_{1}^{2}$ is always negative from the variance expressions in Propositions 2, 4 and 6; and the natural logarithm term is always negative. Hence, we must examine the term $\frac{\hat{\sigma}_{0}^{2} \mu_{1} - \hat{\sigma}_{1}^{2} \mu_{0}}{\hat{\sigma}_{0}^{2} - \hat{\sigma}_{1}^{2}}$. If this term is smaller than the square root term then we are done: the negative solution in (\ref{thm2}) takes on a negative value, which is not feasible with energy detection. If this term is larger than the square root term, then both solutions of (\ref{thm2}) take on a positive value. 

For the latter case, we wish to show that one solution to (\ref{thm2}) is always smaller than $\mu_{0}$; and as a result, the solution closer to the ML threshold is always the positive solution of (\ref{thm2}). Let $\mu_{1}\!=\!(1\!+\!\varepsilon) \mu_{0}$ where $\varepsilon\!>\!0$, and let $\hat{\sigma}_{0}^{2}, \hat{\sigma}_{1}^{2}$ take on arbitrary values. Simplifying gives
\begin{align}
\frac{\hat{\sigma}_{0}^{2} \mu_{1} - \hat{\sigma}_{1}^{2} \mu_{0}}{\hat{\sigma}_{0}^{2} - \hat{\sigma}_{1}^{2}} &= \frac{\hat{\sigma}_{0}^{2} (1 + \varepsilon) \mu_{0} - \hat{\sigma}_{1}^{2} \mu_{0}}{\hat{\sigma}_{0}^{2} - \hat{\sigma}_{1}^{2}} \nonumber \\
	&= \frac{ \mu_{0} \left( \hat{\sigma}_{0}^{2} (1 + \varepsilon) - \hat{\sigma}_{1}^{2} \right) } {\hat{\sigma}_{0}^{2} - \hat{\sigma}_{1}^{2}} \nonumber \\
	&= \mu_{0} \left( 1 + \frac{\varepsilon \hat{\sigma}_{0}^{2}}{\hat{\sigma}_{0}^{2} - \hat{\sigma}_{1}^{2}} \right).	\label{appC_3}
\end{align}
Now we need to show that, $\forall \varepsilon > 0$,
\begin{equation}
\left( 1 + \frac{\varepsilon \hat{\sigma}_{0}^{2}}{\hat{\sigma}_{0}^{2} - \hat{\sigma}_{1}^{2}} \right) \mu_{0} - \sqrt{\frac{\hat{\sigma}_{0}^{2} \hat{\sigma}_{1}^{2} \left( \left(\mu_{0} - \mu_{1}\right)^{2} + 2 \left(\hat{\sigma}_{0}^{2} - \hat{\sigma}_{1}^{2}\right) \ln\left(\frac{\hat{\sigma}_{0}}{\hat{\sigma}_{1}}\right)\right)}{\left(\hat{\sigma}_{0}^{2} - \hat{\sigma}_{1}^{2}\right)^{2}}} < \mu_{0}.	\label{appC_4}
\end{equation}
The square root term in (\ref{appC_4}) can be lower bounded by:
\begin{equation}
\sqrt{\frac{\hat{\sigma}_{0}^{2} \hat{\sigma}_{1}^{2} \left( \left(\mu_{0} - \mu_{1}\right)^{2} + 2 \left(\hat{\sigma}_{0}^{2} - \hat{\sigma}_{1}^{2}\right) \ln\left(\frac{\hat{\sigma}_{0}}{\hat{\sigma}_{1}}\right)\right)}{\left(\hat{\sigma}_{0}^{2} - \hat{\sigma}_{1}^{2}\right)^{2}}} \geq \varepsilon \mu_{0} \sqrt{\frac{\hat{\sigma}_{0}^{2} \hat{\sigma}_{1}^{2}}{\left( \hat{\sigma}_{0}^{2} - \hat{\sigma}_{1}^{2} \right)^{2}}}.	\label{appC_5}
\end{equation}
Substituting (\ref{appC_5}) into (\ref{appC_4}), the condition to prove becomes ($\forall \varepsilon > 0$):
\begin{equation}
\left( 1 + \frac{\varepsilon \hat{\sigma}_{0}^{2}}{\hat{\sigma}_{0}^{2} - \hat{\sigma}_{1}^{2}} \right) \mu_{0} - \varepsilon \mu_{0} \sqrt{\frac{\hat{\sigma}_{0}^{2} \hat{\sigma}_{1}^{2}}{\left( \hat{\sigma}_{0}^{2} - \hat{\sigma}_{1}^{2} \right)^{2}}} < \mu_{0}. 	\label{appC_5a}
\end{equation}
Simplifying and rearranging (\ref{appC_5a}), we have
\begin{equation}
\frac{\hat{\sigma}_{0}^{2}}{\hat{\sigma}_{0}^{2} - \hat{\sigma}_{1}^{2}} < \sqrt{\frac{\hat{\sigma}_{0}^{2} \hat{\sigma}_{1}^{2}}{\left( \hat{\sigma}_{0}^{2} - \hat{\sigma}_{1}^{2} \right)^{2}}}.	\label{appC_6}
\end{equation}
As $\frac{\hat{\sigma}_{0}^{2}}{\hat{\sigma}_{0}^{2} - \hat{\sigma}_{1}^{2}} < 0$, the negative solution to (\ref{thm2}) is always smaller than $\mu_{0}$ regardless of $\varepsilon$. The negative solution is closer to $\frac{\mu_{0} + \mu_{1}}{2}$ (a good approximation for the ML threshold at low SNR, see Fig. 2) when $\mu_{0} = \mu_{1}$. Hence, the positive solution to (\ref{thm2}) is always closer to the ML threshold for all other $\mu_{0}$. When $|B_{0}| > |B_{1}|$, (\ref{appC_6}) no longer holds, as $\hat{\sigma}_{0}^{2} > \hat{\sigma}_{1}^{2}$; and by contradiction, the negative solution must be the one closer to the ML threshold. This completes the proof.

\bibliographystyle{ieeetran}
\bibliography{IEEEabrv,afdf_ref}

\end{document}